\documentclass[12pt, preprint]{emulateapj}
\usepackage{natbib}
\usepackage{subeqnarray}
\usepackage{topcapt}

\newcommand{\eg}{e.g.\ }
\newcommand{\p}{\partial}

\newcommand{\ts}{t_{\rm stop}}
\newcommand{\taus}{\tau_{\rm s}} 
\newcommand{\taue}{\tau_{\rm e}}

\newcommand{\Sc}{\mathrm{Sc}}
\newcommand{\St}{\mathrm{St}}

\DeclareMathSymbol{\varOmega}{\mathord}{letters}{"0A}
\DeclareMathSymbol{\varSigma}{\mathord}{letters}{"06}
\DeclareMathSymbol{\varPsi}{\mathord}{letters}{"09}

\newcommand{\Eq}[1]{equation\,(\ref{#1})}
\newcommand{\Eqs}[2]{equations (\ref{#1}) and~(\ref{#2})}

\newcommand{\Fig}[1]{Fig.~\ref{#1}}

\newcommand{\Tab}[1]{Table \ref{#1}}

\newcommand{\brak}[1]{\langle #1\rangle}

\newcommand{\te}{t_{\rm eddy}}

\newcommand{\up}{u_{\rm p}}
\newcommand{\ug}{u_{\rm g}}
\newcommand{\vp}{v_{\rm p}}
\newcommand{\vg}{v_{\rm g}}
\newcommand{\vzp}{w_{\rm p}}
\newcommand{\wg}{w_{\rm g}}
\newcommand{\xp}{x_{\rm p}}
\newcommand{\yp}{y_{\rm p}}
\newcommand{\zp}{z_{\rm p}}
\newcommand{\gs}{_{\rm g}}
\newcommand{\ps}{_{\rm p}}

\slugcomment{Draft Modified \today}

\shorttitle{Particles in Turbulence}
\shortauthors{Youdin \& Lithwick }

\begin{document}

\title{Particle Stirring in Turbulent Gas Disks: Including Orbital Oscillations}
\author{Andrew N. Youdin\altaffilmark{1} \& Yoram Lithwick}
\affil{C.I.T.A., U. of Toronto, 60 St. George St., Toronto, ON M5S 3H8}
\altaffiltext{1}{youd@cita.utoronto.ca}

\begin{abstract}
We describe the diffusion and random velocities of solid particles  due to stochastic forcing by turbulent gas. We include the orbital dynamics of Keplerian disks, both in-plane epicycles and vertical oscillations.  We obtain a new result for the diffusion of solids.
The Schmidt number (ratio of gas to particle diffusivity) is $\Sc \approx 1 + (\varOmega \ts)^2$, in terms of the particle stopping time $\ts$ and  the orbital frequency $\varOmega$.  The standard result, $\Sc = 1 + \ts/\te$, in terms of the eddy turnover time, $\te$, is shown to be  incorrect.  The main difference is that $\Sc$ rises quadratically, not linearly, with stopping time.  Consequently, particles larger than $\sim 10$ cm in protoplanetary disks will suffer less radial diffusion and will settle closer to the midplane.  Such a layer of boulders would be more prone to gravitational collapse.  Our predictions of RMS speeds, vertical scale height and diffusion coefficients will help interpret numerical simulations.  We confirm previous results for the vertical stirring of particles (scale heights and random velocities), and add a correction for arbitrary ratios of eddy to orbital times.  The particle layer becomes thinner for $\te > 1/\varOmega$ with the strength of turbulent diffusion held fixed.   We use two analytic techniques -- the Hinze-Tchen formalism  and the Fokker-Planck equation with velocity diffusion -- with identical results when the regimes of validity overlap.  We include simple physical arguments for the scaling of our results.
\end{abstract}


\section{Introduction}
The stirring of particles by turbulent gas is a fundamental problem in fluid dynamics.  It is especially important in studying the formation of planetesimals because the (non-Keplerian) motion of particles below about a kilometer in radius is strongly influenced by gas drag.  Our goal here is to include, for the first time, the effect of Keplerian orbital dynamics on the random speeds and diffusion of particles.


Circumstellar disks are thought to be turbulent.  Turbulence drives accretion onto protostars by dissipating Keplerian shear and transferring angular momentum radially outwards.  The magneto-rotational instability \citep{bh91} is a leading candidate for the source of disk turbulence despite concerns that protoplanetary disks may be insufficiently ionized in ``dead zones" near the disk midplane \citep{gam96, swh04}. Convection \citep{lp80, khk99} and purely hydrodynamical mechanisms \citep{dub05,roc07,lith07} to generate turbulence are being actively investigated.  For this work, we remain agnostic as to the source of disk turbulence, as well as to its detailed properties. 

While turbulence is beneficial for accretion onto stars, it is thought to hinder the growth of planetesimals in disks.  The gravitational collapse of smaller solids into planetesimals \citep{saf69,gw73,ys02} is opposed by turbulence \citep{stu80}.  Turbulent diffusion balances the vertical settling of dust grains and larger solids to the midplane (\citealp{cdc93}, hereafter CDC, \citealp{sek98,dms95}, \citealp{cfp06}, hereafter CFP), thereby setting the midplane volume density for a given surface density.  The velocity dispersion of solids determines the \citet{too64} gravitational stability, or $Q$, parameter for thin disks.  By analogy with single component disks, it is argued that gravitational collapse requires densities above a critical threshold, and/or $Q$ values below unity.  This overlooks the important role of gas drag in the collapse of solids, which removes such thresholds \citep{war76,war00,y05a,y05b}.  These works show that collapse into rings over many dynamical times is possible even for nominally gravitationally stable values of midplane density and velocity dispersion.  Radial diffusion plays an important role in this type of dissipative collapse.  This work establishes formulae for vertical density structure, velocity dispersion and radial diffusion of particles in turbulent gas disks.  These results can be used to construct more realistic models of gravitational collapse of solids into planetesimals.

Turbulent stirring also determines the relative speeds with which particles collide \citep[][hereafter VJMR]{vjmr80} and can be used to help determine when particles coagulate, fragment, or inelastically dissipate energy \citep{dt97}.  Since our work does not include the spatial dynamics internal to eddies we do not specifically address relative velocities.  However for large solids with long aerodynamic stopping times the relative speeds are just the RMS random speeds (added in quadrature).    Other works give refinements \citep{mmv91} and analytical fits \citep{oc07} to the VJMR relative velocities, and investigate the implications for coagulation, given various assumptions about particle sticking \citep{mmv88}.  VJMR does not give results for particle diffusion and does not include orbital dynamics.

Recent years have seen a remarkable advance in numerical simulations of the dynamics of solid particles in turbulent gas disks (\citealp[e.g.][]{jk05,fn05,csp05}).  The results of this work can be used to make detailed comparisons to the motion of test particles in these simulations.  Simulations have now begun to include the dynamical feedback of particles on gas dynamics \citep{jhk06, jy07}.  These works find that streaming instabilities \citep{yg05, yj07} lead to the clumping of particles, an effect our turbulent stirring models do not include.  Other particle concentration mechanisms, including long-lived vortices \citep{fmb01}, turbulent pressure enhancements \citep{jkh06}, or low-enstrophy regions at the dissipation scale \citep{hc07} are neglected as well.

The standard result for particle diffusion was given by CDC  (see \S\ref{s:CDC}).  We improve on this result in two ways.  First we evaluate the statistical growth of particle separations, instead of making the (incorrect) assumption that diffusion is proportional to kinetic energy.  Second, we include orbital dynamics, both vertical oscillations and the epicyclic motion that couples radial and azimuthal forcing.  Since the CDC result was correct for small, tightly-coupled solids, many previous results are unaffected by the error.  


This work is organized as follows.  We introduce the relevant timescales of the problem in \S\ref{s:timescale}.  Our main results are justified with simple order of magnitude estimates in \S\ref{s:OOM}.  We review particle stirring with the Hinze-Tchen formalism in \S\ref{s:Hfree}, before extending it to include vertical oscillations in \S\ref{s:Hvert} and epicyclic oscillations in \S\ref{s:Hepi}.  We confirm our results with the Fokker-Planck equation in \S\ref{s:FP}.  Section \ref{s:neglect} addresses neglected effects such as collisions (\S\ref{s:coll}), the ``crossing trajectories" effect (\S\ref{s:CT}), and particle feedback on gas dynamics (\S\ref{s:feedback}).  We discuss the implications of our results in \S\ref{s:Disc}.

\section{Timescales, Dimensionless Parameters, and Particle Sizes}\label{s:timescale}
 Particle stirring in turbulent gas disks is characterized by three timescales: $\ts$, $\te$, and $1/\varOmega$.  The particle stopping time, $\ts$, is the exponential timescale for the decay of relative motion with gas via drag forces, e.g. \Eq{eq:Langevin}, and $\ts$ is an increasing function of particle size (for a given material density) as detailed in appendix \ref{s:sizes}.  The eddy time, $\te$, is the correlation time of turbulent fluctuations (defined formally in equation \ref{eq:Dg}).  For Kolmogorov turbulence, $\te$ is also the turnover time of the largest-scale eddies.  Smaller eddies have shorter turnover times, and contribute less to the turbulent kinetic energy.  The orbital frequency, $\varOmega$, is also the vertical and epicyclic oscillation frequency in Keplerian disks,  where $1/\varOmega$ is also the radial shear time (modulo $3/2$).
 
\begin{figure}[htb] 
\hspace{-1.5cm}    \includegraphics[width=4in]{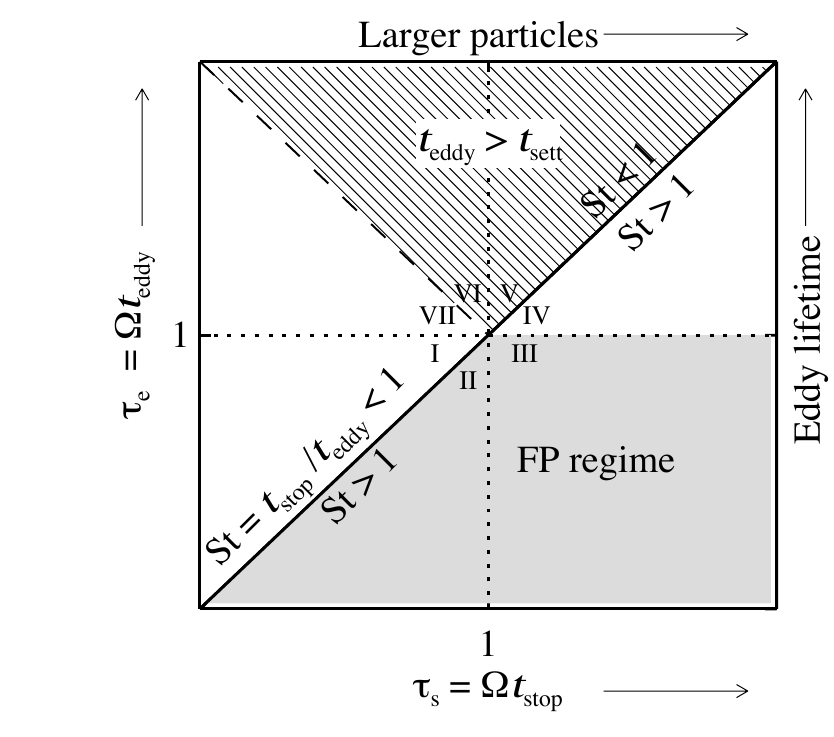} 
    \caption{Schematic plot of the parameter space for particle stirring with eddy time vs.\ particle stopping time, both normalized to the orbital frequency.  Dotted lines indicate where these $\tau_{\rm e}$ and $\taus$ parameters are unity.  The straight line of slope one demarcates Stokes number of unity, where particles are marginally coupled to eddies.  The shaded region indicates where a Fokker-Planck approach is valid because the stochastic forcing time, $\te$, is short.  In the upper hatched region, $\te$ exceeds the vertical settling time.  Roman numerals label regions where timescales are ordered as indicated in \Tab{tab:ordering}
        . }
    \label{fig:tets}
 \end{figure}

We define three dimensionless parameters, of which two are independent,
\begin{subeqnarray} 
 \taus &\equiv& \varOmega \ts \\
\tau_{\rm e} &\equiv& \varOmega \te \\
\St &\equiv&  \ts/\te = \taus/\tau_{\rm e}\, ,
\end{subeqnarray}
to describe the parameter space of turbulent stirring, as shown schematically in \Fig{fig:tets}.  The dimensionless stopping time measures whether particles are tightly ($\taus \ll 1$) or weakly ($\taus\gg 1$) coupled to unperturbed circular gas orbits.  The Stokes number, $\St$, also measures particle coupling, but to turbulent fluctuations. The dimensionless eddy time, $\tau_{\rm e}$, describes the effect of orbital shear on eddies.  If  eddies are destroyed
in a shear time, then $\tau_{\rm e} \sim 1$, and some local MRI simulations support this choice \citep{fp06}.\footnote{Note however that \citet{fp06} use a local Eulerian measure of the eddy time, which they argue should be similar to the more appropriate Lagrangian measure.}   Orbital dynamics plays no role in turbulence with $\tau_{\rm e} \ll 1$, which is unsheared and would require an energy source other than orbital differential rotation, perhaps convection.
Turbulent eddies with $\tau_{\rm e} \gg 1$\footnote{Even though 
long-lived vortices are often invoked to trap
 particles, turbulence with $\tau_{\rm e}\gg 1$ does not necessarily concentrate
 particles within eddies. 
   To trap particles, a vortex must live for many circulation times
  (the time for fluid to circulate
 within a vortex).
For example,   in \S \ref{s:turbmodel} we introduce a simple model for 
 $\tau_{\rm e}\gg 1$ turbulence
 that is composed
 of vortices that live for a single circulation time. 
  Even though their circulation time is much longer than the orbital
 time, 
 these vortices cannot concentrate particles in their lifetime.
  }
  are extended azimuthally by orbital shear.  MRI turbulence might well give rise to $\tau_{\rm e} \gg 1$ eddies as indicated by  the extended azimuthal structure seen in MRI simulations \citep{hgb95} and perhaps particle trapping seen in global MRI simulations \citep{fn05}.

Given our ignorance of the detailed properties of turbulence in protoplanetary disks, we consider deviations from the favored value of $\tau_{\rm e} = 1$  (in which case $\taus  = \St$).  By keeping $\tau_{\rm e}$ general, we also gain a physical understanding of its effect.  Appendix \ref{s:turbmodel} describes a simple model for turbulent velocities with arbitrary $\tau_{\rm e}$, which includes basic dynamical considerations of shearing disks and is useful in making numerical evaluations.
   We include general expressions for our results so other realizations of turbulence can be considered.  
 
Previous works on particle stirring which neglect orbital dynamics effectively assume $\taus, \tau_{\rm e} \ll 1$ (regions I and II in the lower left corner of \Fig{fig:tets}).  Our study of vertical motions in \S\ref{s:Hvert}, and epicyclic oscillations in \S\ref{s:Hepi} includes all regions of  \Fig{fig:tets}.
The shaded region in \Fig{fig:tets} indicates where $\te$ is the shortest timescale in the problem (regions II and III).   Here the Fokker-Planck equation can model turbulent stirring with the inclusion of velocity diffusion, as done here (\S\ref{s:FP}) for in-plane motion and by CFP for the vertical settling problem.   

The  vertical settling timescale, $t_{\rm sett}$, of tightly coupled particles, $\taus \ll 1$,  at the terminal velocity for stellar gravity is $t_{\rm sett} = 1/(\varOmega \taus)$.  The vertical oscillations of loosely coupled particles damp in $t_{\rm sett} = \ts$.  These results can be combined to give:
\begin{equation} \label{eq:tsett}
\varOmega t_{\rm sett}\simeq \taus+1/\taus
\end{equation}  
The hatched triangle in \Fig{fig:tets} (regions V and VI) shows where the eddy times exceed particle settling times.

\subsection{Aside on the Schmidt Number}\label{s:Sc}
This paper obtains a new result for the spatial diffusion of heavy particles, $D\ps$.  We compare this to the spatial diffusion of gaseous contaminants, $D\gs$, which depends only on the time-integrated correlation of turbulent velocities (equation \ref{eq:Dg}).  Following CDC we call the ratio $\Sc \equiv D\gs/D\ps$, the Schmidt number, and we generally expect $\Sc \gtrsim 1$ because particles should not diffuse faster than the gas.  This notation is somewhat unfortunate as it differs from the standard fluid dynamical convention.  In fluid dynamics $\Sc_{\rm hydro} \equiv \nu/D\gs$, is the ratio of the viscous (momentum) diffusion to mass diffusion (of the fluid itself for the typical single fluid case).  This ratio is astrophysically important since it determines the diffusion of gas (or small particle) contaminants relative to the accretion flow \citep{ste90,twby06}.  Most numerical studies of particle diffusion \citep{jk05,csp05} measure the Schmidt number for particles as the ratio $\Sc_{\rm sim} = \nu/D\ps$, where $\nu$ includes the total accretion stress, with Reynolds and Maxwell stress contributions,  since the turbulence is MRI-driven.  Thus $\Sc_{\rm sim}$ combines the effects of particle dynamics and  angular momentum transport.  Indeed, \citet{jkm06} found that imposed vertical magnetic flux significantly affected $\Sc_{\rm sim}$.  This explains why there have been reports of $\Sc_{\rm sim}$ both above and below unity for tightly coupled particles, while we find (with CDC) that $\Sc = 1$ in this limit.
To isolate the effects of particle dynamics, and to compare directly with CDC, we opt for the CDC convention of $\Sc$.  Comparison of our results with numerical simulations requires measurement of both the diffusion of heavy particles, $D\ps$, and Lagrangian tracers of the fluid motion, $D\gs$.

\section{Physical Scaling Arguments}\label{s:OOM}

Our main results can be understood with simple physical arguments in limiting cases.  We assume isotropic turbulence with a RMS speed of $\delta V\gs$.  The turbulent diffusion coefficient for gas is $D\gs \sim \delta V\gs^2\, \te$.

\subsection{No Orbital Oscillations: $\taus \ll 1,\tau_{\rm e}\ll 1$}
We first ignore  orbital  effects by taking $\tau_{\rm e} \ll 1$ and $\taus \ll 1$ (lower left quadrant of \Fig{fig:tets}), so the particle response depends only on $\St$.  For $\St \ll 1$ particles are well coupled to eddies.  Thus the particle RMS velocities, $\delta V\ps \sim \delta V\gs$, and diffusion coefficient, $D\ps \sim D\gs$, are the same as for the turbulent gas.

For $\St \gg 1$  turbulent eddies wash over loosely coupled particles, and the kicks give a random walk in velocity.  Each kick of duration $\te$ has an amplitude $V_{\rm kick} \sim \delta V_{\rm g}~ \te/\ts$ in a random direction.  
Particle motions equilibrate with the turbulent forcing in $\ts$.  Thus after $N = \ts/\te$ kicks, the random velocity will saturate at:
\begin{equation} \label{eq:vpest}
\delta V_{\rm p} \approx \sqrt{N} V_{\rm kick} = \delta V_{\rm g}/\sqrt{\St} \mathrm{~for~} \St \gg 1 \, .
\end{equation}
By combining the large and small $\St$ limits we recover the standard result of \Eq{eq:upsq}.

To estimate particle diffusion  for $\St \gg 1$, note it takes $\ts$ to deviate the particle velocity by order unity.  Thus the particle mean free path is $\ell\ps = \delta V_{\rm p} \, \ts$, and the diffusion is $D_{\rm p} \sim \delta V_{\rm p}\, \ell\ps \sim \delta V_{\rm p}^2 \, \ts \sim \delta V_{\rm g}^2 \, \te \sim D_{\rm g}$.  With increasing $\St$, the longer mean free path cancels the lower velocities to maintain $D\ps = D\gs$ for  all $\St$ in agreement with \Eq{eq:Dp}, which also ignores orbital oscillations.  Analysis of previous results which erroneously claim that $D\ps$ decreases for large $\St$ in this regime can be found in \S\ref{s:CDC}.

\subsection{Including Orbital Oscillations: $\taus \gg 1 \gg \tau_{\rm e}$}
To better understand orbital oscillations we consider $\taus \gg 1$, so that particles decouple from gas orbits, but keep $\tau_{\rm e} \ll 1$.  Thus particles receive many small kicks during a vertical oscillation or an epicycle and $\St \gg 1$.   This region occupies the lower right quadrant of \Fig{fig:tets}.  The particle velocity response is still a random walk, which saturates at $\delta V\ps \sim \delta V\gs/\sqrt{\St}$  as above, unaffected by oscillations with frequency $\varOmega$.  This is confirmed in \Eqs{eq:wpsq}{eq:shorteddy}, where the latter includes corrections for the ratio of radial to azimuthal epicyclic motion.

Spatial particle diffusion is interrupted by epicyclic oscillations with a length scale $\ell_{\rm epi} \sim \delta V_{\rm p}/\varOmega$.   
Vertical oscillations are centered on the midplane, and thus do not freely diffuse over long times.   For epicyclic motion, we estimate diffusion by noting that the guiding center shifts by $\ell_{\rm epi}$ every $\ts$, because the velocity changes by $\delta V\ps$ in this same interval.  This gives a diffusion coefficient:
\begin{equation}\label{eq:Dpest}
D_{\rm p} \sim {\ell_{\rm epi}^2 \over \ts} \sim {\delta V_{\rm p}^2 \over \varOmega \taus}  \sim {\delta V_{\rm g}^2 \te \over  \taus^2} \sim {D_{\rm g} \over \taus^2}\, .
\end{equation} 
This can be confirmed by considering the more incremental changes during each orbital oscillation.  Every $1/\varOmega$ a particle receives $N_{\rm epi} = 1/\tau_{\rm e}$ velocity kicks of $V_{\rm kick} \sim \delta V\gs/\St$ to give a velocity change, $\delta V_{\rm epi} \sim V_{\rm kick}\sqrt{N_{\rm epi}}\sim \delta V\gs\sqrt{\tau_{\rm e}}/\taus$, every orbital period.  Thus every orbital period gives a shift in the guiding center position of $\delta \ell_{\rm epi} \sim \delta V_{\rm epi}/\varOmega$.  A random walk with steps of size $\delta \ell_{\rm epi}$ every orbital time gives a particle diffusion coefficient $D_{\rm p} \sim {\delta \ell_{\rm epi}^2 \varOmega} \sim D\gs/\taus^2$ as in \Eq{eq:Dpest}.

If we include the $\taus \ll 1$ limit where orbital effects are negligible, then our estimate of the Schmidt number is:
\begin{equation} 
\Sc  \equiv {D_{\rm g} \over D\ps} \sim 1 +  \taus^2 \label{eq:diskest}
\end{equation}
To order unity, this agrees with the precise result for radial diffusion, \Eq{eq:Scx}, which holds for all $\tau_{\rm e}$.  Thus orbital oscillations reduce the diffusion of loosely coupled particles. 

\subsection{Vertical Scale Height}\label{s:vertest}

We obtain the particle scale height, $H\ps$, by equating the diffusion time, $H\ps^2/D\ps$, and the settling time (\Eqs{eq:tsett}{eq:diskest}). The result,
\begin{equation} \label{eq:Hpapp}
H\ps \sim \sqrt{D\gs\over \varOmega \taus}\, ,
\end{equation} 
holds for arbitrary $\taus$ and agrees with our detailed results in \Eq{eq:Hpsq} and previous works.  In the loose coupling limit, $H\ps$ can be obtained simply by noting that $H\ps \sim \delta V\ps/\varOmega$ for vertical oscillations.  CFP computed $H\ps$ in the loose coupling limit both analytically and with hydrodynamic simulations and found \Eq{eq:Hpapp}.\footnote{CFP also show that both the spatial and velocity distributions are Gaussian with the stated RMS centered on zero velocity at the midplane.}      The simple link between \Eqs{eq:diskest}{eq:Hpapp} confirms that orbital oscillations weaken particle diffusion.

 \begin{table*}[t]
    \centering
    \begin{minipage}{160mm}
    \caption{Summary of Results and Ordering of Timescales by Region in \Fig{fig:tets}.} 
\begin{center} \begin{tabular}{cc|cccc} 
 Region & Timescales &${D_{\mathrm{p},x} / D_{\mathrm{g},x}}$\tablenotemark{a} &${H\ps/ H\gs}$\tablenotemark{b} &${\delta u\ps/ \delta u\gs}$\tablenotemark{c} &${\delta w\ps/\delta w\gs}$\tablenotemark{c} 
\\
\hline
I & $\taus<\taue<1$ &1& $\sqrt{\alpha_z / \taus}$
&1&$1$
 \\
II & $\taue<\taus<1$ &1& $\sqrt{\alpha_z / \taus}$
&$\sqrt{\taue/\taus}$ &$\sqrt{\taue/\taus}$

 \\
III & $\taue<1<\taus$ &$4/\taus^2$ & $\sqrt{\alpha_z / \taus}$
&$\sqrt{5 \taue/2\taus}$& $\sqrt{\taue/\taus}$
 \\
IV & $1<\taue<\taus$ &$4/\taus^2$& $\sqrt{\alpha_z / \taus\taue^2}$
&$\sqrt{5/2\taus \taue}$& $1/\sqrt{\tau_s\tau_e}$
 \\
V & $1<\taus<\taue$ &$4/\taus^2$& $\sqrt{\alpha_z /\taus^2 \taue}$
&$2/\taus$\tablenotemark{d}& $1/\sqrt{\tau_s\tau_e}$\tablenotemark{d}
 \\
VI & $\taus<1<\taus^{-1}<\taue$ &1& $\sqrt{\alpha_z / \taus^2 \taue}$
&$1$\tablenotemark{d}& $1/\sqrt{\tau_s\tau_e}$\tablenotemark{d}
 \\
VII &$\taus<1<\taue<\taus^{-1}$  &1& $\sqrt{\alpha_z / \taus}$
&1& 1
\end{tabular}\end{center}
\tablecomments{Results for particle radial diffusion ($D_{\mathrm{p},x}$), scale height ($H\ps$) and radial ($\delta u\ps$) and azimuthal ($\delta w\ps$) RMS velocities from eqs.\ (\ref{eq:Dpx}), (\ref{eq:Hpsq}), (\ref{eq:episq}a), and (\ref{eq:wpsq}) in the appropriate limiting cases.  The in-plane results (diffusion and radial velocity) assume isotropic radial and azimuthal speeds.}
\tablenotetext{a}{Regions II, III, IV and V differ from the previous result for $D_{\rm p}$ in eq.\ (\ref{eq:ScCDC}) which gave $1/\St$ for II, III, and IV and 1 for region V.}
\tablenotetext{b}{Regions IV, V and VI all give a thinner particle scale height than the standard result $\sqrt{\alpha_z / \taus}$ in eq.\ (\ref{eq:Hpstd}).}
\tablenotetext{c}{Regions IV and V for radial speeds and Regions IV, V and VI for vertical speeds (all of which have long eddy times) differ from the standard result $1/\sqrt{1 + \St}$ from eq.\ (\ref{eq:upsq}).}
\tablenotetext{d}{The radial and azimuthal speeds have a different scaling in regions V and VI because the eddy time is longer than the vertical settling time.}
    \label{tab:ordering}
    \end{minipage}
 \end{table*}

\section{Fourier Solution of Langevin Equations}\label{s:HT}
\subsection{Hinze-Tchen Formalism: No Orbital Oscillations}\label{s:Hfree}
We summarize the classic derivation of \citet{Hinze} and \citet{Tchen} for turbulent transport of particles in the absence of orbital effects or other external forces \citep[reviewed in][]{fanzhu}.   
A Langevin equation describes the forcing of particle velocities, $\up$, by a gas velocity, $\ug$, with stochastic fluctuations as
\begin{equation}\label{eq:Langevin}
{d \up \over dt} = -{\up \over \ts} + {\ug(t) \over \ts}\, .
\end{equation}
We restrict our attention to a single dimension with no loss of generality since the motions 
in different dimensions
are uncoupled.  The drag force is linear in the relative velocity, which applies in either the Epstein or Stokes regimes.  For protoplanetary disks, this requires particle sizes below $a_{\rm turb}$ in \Eq{eq:aturb}.  Spatial  variations in the gas forcing are ignored, because the time to cross an eddy is longer than the eddy lifetime (see \S\ref{s:CT}).

We decompose $\up$ and $\ug$ into Fourier modes that vary as $\exp(-\imath \omega t)$.  The  Fourier amplitudes evolve by \Eq{eq:Langevin} as:
\begin{equation} 
\hat{u}\ps = {\hat{u}\gs \over 1-\imath \omega \ts}
\end{equation}
The power spectrum of turbulent forcing,
\begin{equation} \label{eq:power}
\hat{E}_{\rm g}(\omega) \equiv \brak{\ug^2}\hat{P}(\omega) =  {\brak{\ug^2} \over \pi} {\te \over 1 + \omega^2\te^2}\, 
\end{equation}
is normalized to the expectation value of turbulent velocity squared, 
\begin{equation} \label{eq:ugsq}
\brak{\ug^2} = \int_{-\infty}^\infty d\omega \hat{E}\gs \, .
\end{equation}  
The frequency dependence in \Eq{eq:power} corresponds to exponential decay of time correlations,  as the Fourier transform shows, $P(t) = \exp(-|t|/\te)/(2\pi)$.  This spectrum is Kolmogorov since $\hat{P}(\omega) \propto \omega^{-2}$ for frequencies above $1/\te$, the integral scale. This equates to the familiar spatial spectra of $\hat{P}(k) \propto k^{-5/3}$ since $k v_k^3 = $ constant, and $v_k = \omega/k$ readily gives $k^{-5/3}dk \propto \omega^{-2}d\omega$.

Diffusion of tracer particles (i.e.\ gas) with position, $x\gs(t) = \int_0^t u\gs(\tau)d\tau  + x\gs(0)$,  is given by the statistically averaged growth rate of squared displacements over long times as
\begin{subeqnarray}\label{eq:Dg} 
D\gs &\equiv& {1 \over 2} {d \brak{x\gs^2} \over dt} = \int_0^\infty \brak{\ug(\tau)\ug(0)}d\tau \\
&=& \int_0^\infty d \tau\int_{-\infty}^{\infty} d\omega \hat{E}\gs(\omega)e^{(-\imath \omega \tau)} \\
&=& \pi \hat{E}\gs(0) = \brak{u\gs^2}\te \, 
\label{eq:zerofreq} \end{subeqnarray}\\
where eq.\ (\ref{eq:Dg}b) equates the velocity autocorrelation function to the Fourier transform of the power spectrum, and eq.\ (\ref{eq:Dg}c) exploits that the power spectrum is even in $\omega$ (equivalent to the time invariance of the correlation function).  The final result is the expected product of RMS velocity $\delta V\gs = \brak{u\gs^2}^{1/2}$ and eddy length, $\ell_{\rm eddy} = \delta V\gs \, \te$.


The power spectrum of particle motions 
\begin{equation} \label{eq:powerp}
\hat{E}_{\rm p}(\omega) = \hat{E}\gs (\omega){|\hat{u}\ps|^2 \over |\hat{u}\gs|^2} = {\hat{E}\gs(\omega) \over 1 + (\omega \ts)^2}
\end{equation}
readily produces the mean squared particle velocities
\begin{equation} 
\brak{\up^2} = \int_{-\infty}^{\infty}d\omega \hat{E}\ps(\omega) ={ \brak{\ug^2} \te \over \te + \ts} = {\brak{\ug^2} \over 1+\St} \label{eq:upsq} 
\end{equation}
This agrees with our physical estimate in \Eq{eq:vpest}, and more importantly, with a detailed treatment that includes the spatial spectrum of eddies (VJMR).

Spatial particle diffusion is computed as in \Eq{eq:Dg}.  Since the power spectrum in \Eqs{eq:power}{eq:powerp} is also even in $\omega$, we simply have
\begin{equation}
D\ps  = \pi \hat{E}\ps(0) =  \brak{\ug^2} \te = D\gs \label{eq:Dp}\, ,
\end{equation} 
independent of stopping time. The Schmidt number is
$ \Sc \equiv D\gs/D\ps = 1$
in this regime, as explained in \S\ref{s:OOM} by the longer mean free path for larger $\ts$.

\subsubsection{Comparison with CDC}\label{s:CDC}
The most often cited result for the Schmidt number in the astrophysical literature is from CDC \citep[]{cdc93}.  Their analysis begins exactly as above (\S\ref{s:Hfree}):  using the Hinze-Tchen formalism with the same temporal power spectrum and ignoring orbital dynamics (see their appendix B).   Instead of calculating diffusion from velocity correlations as in \Eqs{eq:Dg}{eq:Dp}, they assume that diffusion coefficients are proportional to the ratio of fluctuating kinetic energies.  This gives,\footnote{CDC further add a correction for the ``crossing trajectories effect" of particles drifting through eddies with a speed $\Delta V$ to get their final result of: $\Sc_{\rm CDC}' = (1+\St)(1+ \Delta V^2/\brak{ u\gs^2})^{1/2}$.  This further increases $\Sc$ and does not explain the discrepancy.  See \S\ref{s:CT} for further discussion.}
\begin{equation} 
\Sc_{\rm CDC} =\brak{\ug^2}/\brak{\up^2} = 1+\St \label{eq:ScCDC}\,  ,
\end{equation} 
which is incorrect since it ignores the role of the particle mean free path.  There is no disagreement for low Stokes numbers where $\Sc \simeq 1$.

\begin{figure}[htbp] 
\hspace{-1.1cm}   \includegraphics[width=4in]{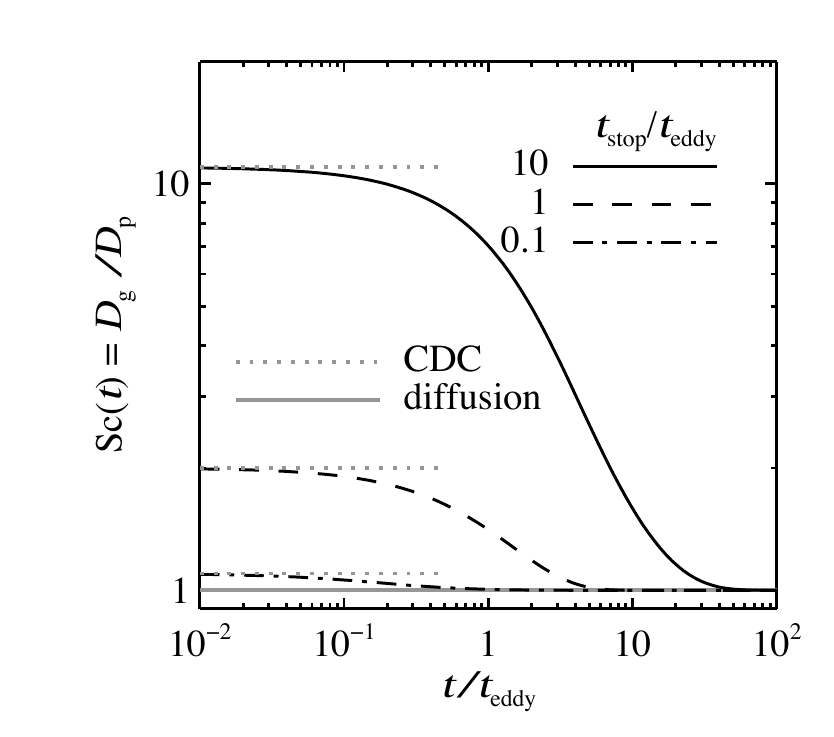} 
   \caption{Schmidt number, the ratio of gas to particle mass transport coefficients, versus time \emph{in the absence of orbital dynamics} or other external forces.  Different black curves choose different values of $\St = \ts/\te$.  In the diffusion limit of long times, $\Sc = 1$ for all $\St$ (\emph{grey sold line}).  The \citet{cdc93} result (\emph{grey dotted line segments}) is valid for $t$ small compared to the eddy and stopping times, but does not describe diffusion over long times.}
   \label{fig:Scvst}
\end{figure}

To better understand the CDC result for $\Sc$, and why it appears in the geophysical literature as well, we define diffusion coefficients for arbitrary time, $D_{\rm p,g}(t) = \int_0^\tau\brak{u_{\rm p,g}(\tau)u_{\rm p,g}(0)}d\tau$, which evaluate to (see also \citealp{fanzhu}):
\begin{eqnarray} 
D\ps(t) &=& { \brak{\ug^2} \te \over 
 \ts^2 - \te^2}
\Big(
\ts^2[1-\exp(-t/\ts)]-  \nonumber \\
&&\te^2[1-\exp(-t/\te)] \Big) \\
D\gs (t)& =& \brak{\ug^2}\te[1-\exp(-t/\te)] \label{eq:Dgt}
\end{eqnarray} 
The short time limit, $t\ll \ts,\te$, recovers the CDC result while the long time limit produces the diffusion result, as demonstrated graphically in \Fig{fig:Scvst}.  The short time limit is of interest in laboratory experiments and in some geophysical contexts for the dynamics internal to the largest eddies.  Astrophysicists however are normally interested in the evolution of protoplanetary disks over many eddy times.  
 Furthermore, the early phase of turbulent mixing does not describe diffusion \emph{per se} since separations grow linearly in time, not as $\sqrt{t}$.

\citet[][]{sh04} reproduce the Schmidt number result of CDC by alternate methods, specifically a mean field theory approach, and they also neglect orbital dynamics.  The source of the discrepancy with our result is more subtle in this case, but it comes down to their choice of a (second order) closure approximation.  A more accurate modeling of the correlation time with their approach should give agreement with our work (Schr\"apler, personal communication).  In any event simple physical arguments should remove any doubt that our result (actually the classic Hinze-Tchen result) gives the correct particle diffusivity for the limiting case of no orbital dynamics.  

Another result for Schmidt number, $\Sc_{\rm Saf} = (1 + \St)^2$ has been attributed to \citet[][hereafter S69]{saf69}, originally by CDC.  
We find no such claim in S69.
CDC's attribution of this result is apparently due to an extension of
S69's theory, which itself
contains errors.  Specifically, S69 gives $\delta V_{\rm p} \approx \delta V_{\rm g}/(1+ \St)$, which is incorrect.  If this is inserted in  \Eq{eq:ScCDC}, it yields the aforementioned expression
for $\Sc_{\rm Saf}$.  The apparent similarity  to our result with orbital effects [\Eq{eq:diskest} with $\taus$ and $\St$ switched] is illusory.


We are thus fully confident that turbulent particle diffusion is independent of $\St$ in the absence of orbital effects, and for the simple turbulence models considered.   This comparison is intended to give confidence in our methods and clarify a confusing field, not to denigrate these important works.  


\subsection{Vertical Oscillations}\label{s:Hvert}
\begin{figure}[htbp] 
   \hspace{-1cm}   \includegraphics[width=4in]{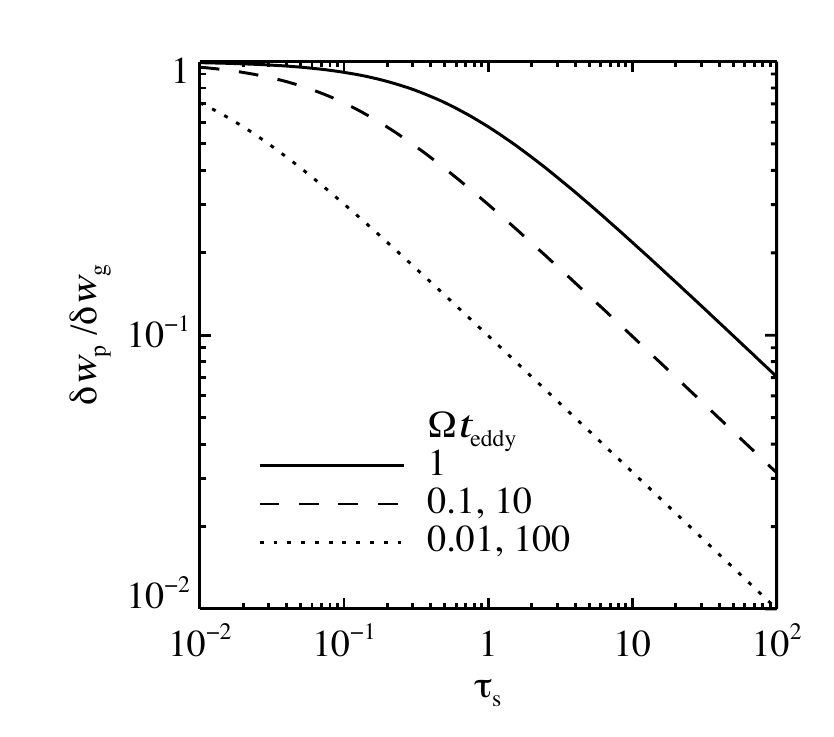} 
   \caption{ Vertical RMS speeds of particles, relative to gas turbulence, vs.\ dimensionless particle stopping time for several choices of the eddy turnover time, from \Eq{eq:wpsq}.  Stirring is most effective for $\taus \ll 1$ and  $\tau_{\rm e} \approx 1$.   For large $\taus$ and/or smaller $\tau_{\rm e}$, particles decouple from eddies.  At large $\tau_{\rm e}$ rapid orbital oscillations cancel the kick from eddies, see \S\ref{s:bigtaue}. }
   \label{fig:wpts}
\end{figure}

We now modify the Langevin equation to include the vertical component of the Keplerian gravitational field:
\begin{equation} \label{eq:vertLangevin}
{d \vzp \over dt} = -\varOmega^2 \zp + {\wg(t) - \vzp \over \ts}
\end{equation} 
where $\vzp = d\zp /dt$ is the vertical particle velocity and $\wg$ the vertical turbulent forcing, which obeys the power spectrum of \Eq{eq:power}.\footnote{With $\brak{w\gs^2}$ instead of $\brak{u\gs^2}$.}  Equation (\ref{eq:vertLangevin}) describes a damped oscillator that is stochastically forced.  The Fourier response of the particle velocity

\begin{equation} 
\hat{w}\ps = {\omega \over \omega + \imath \ts (\varOmega^2 - \omega^2)}\hat{\wg}
\end{equation}
gives the squared random velocities as:
\begin{eqnarray} 
\brak{\vzp^2} = \int_{-\infty}^\infty d\omega \hat{E}\gs(\omega){|\hat{\vzp}|^2 \over |\hat{\wg}|^2}
 &=& {\brak{\wg^2} \over 1 + \taus/\tau_{\rm e}+\taus \tau_{\rm e}} 
 \nonumber
\\
&=& {\brak{\wg^2} \over 1 + \St(1+\tau_{\rm e}^2)}\, ,
\label{eq:wpsq}
\end{eqnarray}
which reduces to the $\varOmega = 0$ result of  \Eq{eq:upsq} for eddies much faster than the orbital time.   \Fig{fig:wpts} plots the vertical RMS particle velocities from the square root of \Eq{eq:wpsq}.  A physical explanation for the large $\tau_{\rm e}$ behavior follows in \S\ref{s:bigtaue}.  The limiting forms of \Eq{eq:wpsq} (as well as eq.\ [\ref{eq:Hpsq}] below) are presented in \Tab{tab:ordering} for each region in \Fig{fig:tets}.


\begin{figure}[htbp] 
\hspace{-1cm}   \includegraphics[width=4in]{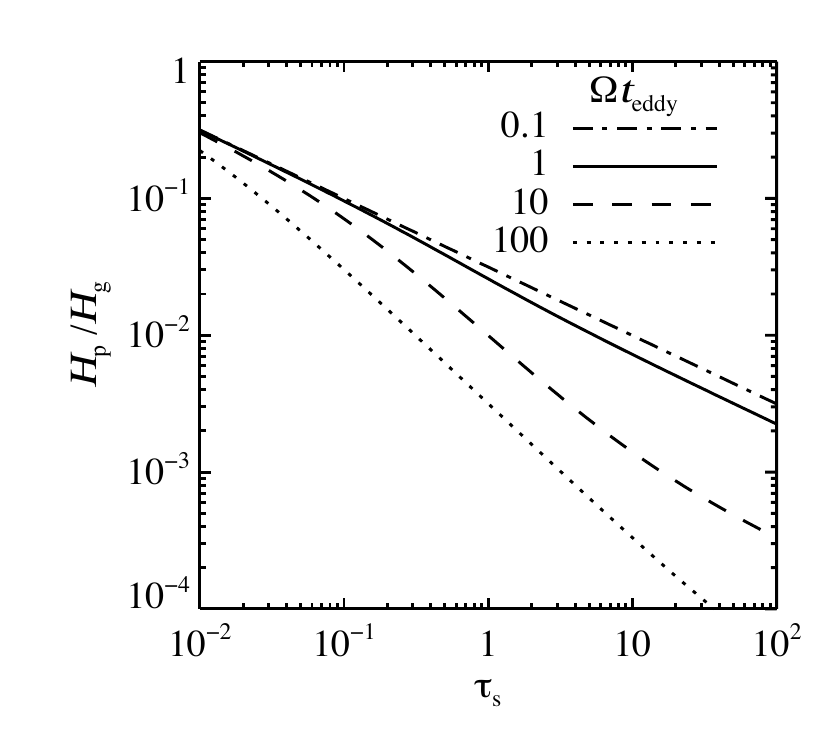} 
   \caption{Particle scale height relative to gas vs.\ stopping time for several values of the eddy time with turbulent diffusion at a level of $\alpha_z \equiv D_{\mathrm{g},z} /(H\gs^2 \varOmega) = 10^{-3}$.  The particle layer becomes thinner for increasing particle stopping time.  The layer also becomes thinner for eddy times longer than the orbital time and the settling time, $1/(\varOmega \taus)$, due to the cancellation of kicks as in \Fig{fig:wpts}.}
   \label{fig:Hpts}
\end{figure}

The particle scale height, $H\ps$,  plotted in \Fig{fig:Hpts}, is calculated using $\hat{z}\ps = \imath \hat{w}\ps/\omega$ as:
\begin{eqnarray}
H\ps^2 &=& \int_{-\infty}^{\infty}d\omega |\hat{z}\ps|^2 {\hat{E}\gs(\omega) \over |\hat{\wg}|^2} =  {D_{\mathrm{g},z} \over \varOmega \taus}{1 \over \xi(\taus,\taue)}  \label{eq:Hpsq} \\ 
\xi &\equiv&  1+{ \taus \tau_{\rm e}^2 \over \taus+\tau_{\rm e}} = 1+ {\St \tau_{\rm e}^2 \over 1+\St}\label{eq:xi}
\end{eqnarray}
using $D_{\mathrm{g},z} = \brak{w\gs^2}\te$.  We compare our result with the standard value derived by \citet{dms95} for tightly coupled particles and by CFP for loose coupling
\begin{equation} 
H_{\rm p}^{(\rm std)} = \sqrt{D_{\mathrm{g},z} \over \varOmega \taus} = \sqrt{\alpha_z \over \taus} H\gs\, , \label{eq:Hpstd}
\end{equation} 
where $\alpha_z \equiv D_{\mathrm{g},z}/(H\gs^2\varOmega)$ is a dimensionless parameter for turbulent diffusion,\footnote{Note that this is not the same $\alpha$-parameter used for diffusion of angular momentum, and that turbulence is subsonic for $\alpha_z /\tau_{\rm e} = \brak{w\gs^2}/c\gs^2 < 1$.} and $H\gs = c\gs/\varOmega$ is the gas scale height. 
Equation (\ref{eq:Hpsq}) reduces to \Eq{eq:Hpstd} when $\xi \rightarrow 1$, i.e.\ for  $\tau_{\rm e} \ll {\rm Max}(1,1/\taus)$ as in regions I, II, III and VII of \Fig{fig:tets}.   A physical description of the large eddy time behavior follows.   For $\tau_{\rm e} = 1$ our result becomes:
\begin{equation} 
H\ps^{\tau_{\rm e} = 1} =  \sqrt{D_{\mathrm{g},z} \over \varOmega \taus} \sqrt{1+\taus \over 1+2\taus}\, .
\end{equation}
The maximum discrepancy of $\sqrt{2}$ for $\taus \gg 1$ is not very significant.
 
Since particles are confined to a vertical potential well, we cannot measure long term spatial diffusion as in \Eq{eq:Dg}.   We show in \S\ref{s:vertest} that the particle diffusivity required to maintain $H\ps$  agrees with the physical estimate of $\Sc$ in \Eq{eq:diskest}.


\subsubsection{Applications and Extensions} \label{s:vertapp}
The ratio between velocity dispersion and scale height,
\begin{equation} 
{\brak{\vzp^2}^{1/2} \over H\ps} = \varOmega \sqrt{\St \over{1+\St}}\, ,
\end{equation} 
reduces to the Keplerian oscillation frequency for loose coupling to eddies, $\St \gg 1$.  One might have naively guessed that the criterion would be $\taus \gg 1$, but see \S\ref{s:OOM} and \S\ref{s:bigtaue} for explanations of the surprising behavior in regions II and V, respectively.   For $\St \ll 1$, collision rates will fall below the usual disk estimate of $\varOmega$ times the vertical optical depth, $\tau\ps$ by $\sqrt{\St}$, since the decrease in density (with larger $H\ps$) is not matched by the increase in RMS speeds.  Since particle relative speeds are reduced below RMS speeds by a factor $\sqrt{\St}$  for $\St \ll 1$ (VJMR) the collision rate is
\begin{equation} 
t_{\rm coll}^{-1} \sim \Omega \tau_{\rm p} {\St \over 1 + \St}\, 
\end{equation}
for equal size particles, assuming that vertical speeds are at least comparable to in-plane motions.  Note that this result is independent of the strength of turbulence (though for different particle sizes turbulence must compete with differential aerodynamic drift).

We also compare the vertical settling speed for $\taus \ll 1$: $w_{\rm sett} = |g_z| \ts \approx \varOmega \taus H\ps$ to RMS particle speeds:
\begin{equation} \label{eq:settRMS}
{w_{\rm sett} \over \brak{\vzp^2}^{1/2}} \approx  \sqrt{(\tau_{\rm e} + \taus)\taus}\, .
\end{equation} 
For $\taus \ll 1$, the above is a small quantity (except for the extreme case of region VI, where $\te > t_{\rm sett}$, see \S\ref{s:bigtaue}), so that random motions exceed the ordered settling that would occur without turbulence.

We have assumed that both the particle stopping time, and the turbulent spectrum are independent of height above the midplane.  This restricts the range of validity of our results to $H\ps < H\gs$ so that gas density, $\rho_{\rm g}$, is roughly constant.   In stratified models, $H\ps \leq H\gs$ is imposed by the hypothesis that the quantity diffused by turbulence is not the absolute particle density, but the particle concentration relative to gas\footnote{See also \citet{ch87}, p. 90, for the same approximation in planetary atmospheres.  The assumption is that turbulent diffusion mimics molecular diffusion, which is driven by concentration gradients (and also thermal gradients, see \citealp{ll59}, eq.\ 58.14).}  \citep[][their eq. 28]{dms95}.   We suggest
\begin{equation} \label{eq:Hpstrong}
{H\ps \over H\gs} \approx \sqrt{\alpha_z \over \alpha_z + \taus}\xi^{-1/2} < 1
\end{equation} 
as a simple way to enforce the restriction $H\ps < H\gs$.

The approximation (that turbulence acts to smooth concentration gradients) does not hold  for larger particles, which move through the gas.  In practice this distinction is not important since $H\ps/H\gs \approx \sqrt{\alpha/\taus} \ll 1$ for $\taus \gg 1$ and $\alpha\lesssim 1$.  Thus heavy particles always settle within $H\gs$ in any event.  See \S\ref{s:FP} for more on the diffusion of heavy particles.


\subsubsection{Physical Understanding of Vertical Forcing for Large $\tau_{\rm e}$}\label{s:bigtaue}
In \S\ref{s:OOM} we provide physical explanations for the velocity dispersion, diffusion, and scale height of particles in the $\tau_{\rm e} \ll 1$ limit, i.e. the lower half of \Fig{fig:tets}.  We now consider $\tau_{\rm e} \gg 1$, which is considerably more complex, particularly in regions V and VI where $\te > t_{\rm sett}$.  We denote the particle and gas RMS speeds as $\delta w\ps$ and $\delta w\gs$, respectively.

We first note that region VII behaves just as region I since particles are tightly coupled to both orbits and eddies, and the eddy time is shorter than the particle settling time.  Thus $\delta w\ps \sim \delta w\gs$ and $H\ps \sim \sqrt{D\gs/(\varOmega \taus)}$ in agreement with \Eqs{eq:wpsq}{eq:Hpsq}, in the limits $\St \ll 1$ and $\St\tau_{\rm e}^2 = \te/t_{\rm sett} \ll 1$.


Region IV has loose particle coupling to eddies and gas orbits, and long eddy times.  As in regions II and III, the velocity grows by a random walk.  However since $\te \gg \varOmega^{-1}$, the kick from an eddy is largely cancelled by the many vertical oscillations in an eddy time.  Thus the kick is only received over the last orbital cycle,   $V_{\rm kick} \sim \delta w\gs/\taus$.  After $N = \ts/\te$ kicks the velocity saturates with $\delta w\ps \sim V_{\rm kick}\sqrt{N} =\delta w\gs/\sqrt{\taus \tau_{\rm e}}$.  Vertical oscillations with loose orbital coupling then gives $H\ps = \delta w\ps/\varOmega$.  Equations (\ref{eq:wpsq},\ref{eq:Hpsq}) agree with these estimates in the relevant $\St \gg 1$, $\St \, \tau_{\rm e}^2 = \St\, \taus \gg 1$ limit.  As a consistency check, note that $H\ps/\ell_{\rm eddy} \sim 1/\sqrt{\taus \tau_{\rm e}^3}\ll 1$ so particle oscillations easily fit within a single large eddy.

Since eddy times exceed settling times in regions V and VI, $H\ps$ is not determined by the usual diffusion arguments.  Instead long-lived eddies suspend particles at $H\ps \sim \delta w\gs/(\varOmega \taus)$, by force balance between gravity, $\varOmega^2 H\ps$, and drag, $\delta w\gs/\ts$, in agreement with \Eq{eq:Hpsq} for $\St \ll 1$ and $\taus^2/\St = \taus \tau_{\rm e} \gg 1$ (regions V and VI).  \citet{hb06} give similar arguments, but without the caveats on the ordering of timescales.  Though ``pinned" by slow eddies, particles will be stirred by faster eddies with  $\te' \sim t_{\rm sett} \ll \te$, and a speed $\delta w\gs' \sim \delta w\gs\sqrt{\te'/\te}$ (the transformation of the Kolmogorov relation $v_\ell \propto \ell^{1/3}$ to the time domain).  For $\taus \ll 1$ (region VI), $\te' \sim t_{\rm sett} =  (\varOmega \taus)^{-1}$, and tight coupling to these slower eddies gives  $\delta w\ps \sim \delta w\gs' \sim \delta w\gs/\sqrt{\taus \tau_{\rm e}}$.   For $\taus \gg 1 $ (region V), $\te' \sim t_{\rm sett} =  \ts$, and the relevant eddy speed is $\delta w\gs' \sim \delta w\gs\sqrt{\St}$.  However the fast oscillations give an effective kick duration of $\varOmega^{-1}$ as in region IV, and $\delta w\ps \sim \delta w\gs'/\taus \sim \delta w\gs/\sqrt{\taus\tau_{\rm e}}$.  There is no accumulation of kicks since  each eddy lasts a stopping time, $\te'\sim \ts$.  Equation (\ref{eq:wpsq}) confirms,  in the 
$\St = \taus/\tau_{\rm e} \ll 1$ and $\taus \tau_{\rm e} \gg 1$ limits, that $\delta w\ps  \sim \delta w\gs /\sqrt{\taus \tau_{\rm e}}$ in regions V and VI as shown in \Tab{tab:ordering}.


These physical arguments justify our new results for the scaling of particle scale height and vertical RMS speeds in the large $\tau_{\rm e}$ regime.

\subsection{Epicyclic Oscillations}\label{s:Hepi}
We now treat the turbulent forcing of epicyclic oscillations.  For small eccentricities, we consider the locally Cartesian radial, $\xp$, and azimuthal, $\yp$, positions relative to a guiding center with Keplerian orbital frequency, $\varOmega$.  Hill's linearized equations of motion with gas drag read: 
\begin{subeqnarray}\label{eq:Hills}
 \dot{x}\ps &=& \up\\
\dot{y}\ps &=&  \vp - (3/2)\varOmega \xp\\
\dot{u}\ps &=&2 \varOmega \vp + {\ug(t)-\up \over \ts}  \\
\dot{v}\ps &=& -{\varOmega \up \over 2} + {\vg(t)-\vp \over \ts} \, ,
\end{subeqnarray}
where we measure azimuthal velocities, $\vp$, relative to local Keplerian shear.  We ignore any time-steady gas velocities, such as pressure-supported sub-Keplerian rotation of the gas would provide.\footnote{The resulting (locally constant) drift of particles relative to gas can still be included with a correction to the eddy time from \Eq{eq:CTE}.}

We assign the same temporal turbulent spectrum, $\hat{P}(\omega)$ as \Eq{eq:power}.  
For generality we will keep the amplitudes of the radial, azimuthal, and correlated gas forcing; $\brak{\ug^2}$, $\brak{\vg^2}$, and $\brak{\ug\vg}$; independent in many expressions to allow comparison with numerical simulations.  \Eq{eq:turbmodel} gives a model to relate these quantities.

\subsubsection{Velocity Response}
The Fourier responses of particle velocities to turbulent forcing are:
\begin{subeqnarray} 
\hat{u}\ps &=& {(1-\imath\nu_{\rm s})\hat{u}\gs + 2\taus \hat{v}\gs \over (1-\imath\nu_{\rm s})^2 + \taus^2} \\
\hat{v}\ps &=& {-(\taus /2)\hat{u}\gs + (1-\imath\nu_{\rm s})\hat{v}\gs  \over (1-\imath\nu_{\rm s})^2 +\taus^2}\, ,
\end{subeqnarray} 
where $\nu_{\rm s} \equiv \omega \ts$.  The squared Fourier amplitudes and correlations read:
\begin{subeqnarray}\label{eq:epiamp}
|\hat{u}\ps|^2 &=& {1\over d(\omega)}\Big[\left(1+\nu_{\rm s}^2\right)|\hat{u}\gs|^2  \nonumber \\
&& 4 \taus^2 |\hat{v}\gs|^2 + 4 \taus {\hat{u}\gs \hat{v}^*\gs + \hat{u}\gs^* \hat{v}\gs \over 2}\Big]\label{eq:epiusq} \\
|\hat{v}\ps|^2 &=&{1\over d(\omega)}\Big[ {\taus^2 \over 4} |\hat{u}\gs|^2+ \left(1+\nu_{\rm s}^2\right)|\hat{v}\gs|^2  -\nonumber\\
&& \taus {\hat{u}\gs \hat{v}^*\gs + \hat{u}\gs^* \hat{v}\gs \over 2}\Big] \label{eq:epivsq}
\\
 {\hat{u}\ps \hat{v}^*\ps + \hat{u}\ps^* \hat{v}\ps \over 2} &=& {1\over d(\omega)}\Big[-{\taus \over 2}|\hat{u}\gs|^2+ 2\taus |\hat{v}\gs|^2  + \nonumber\\
 && \left(1 + \nu_{\rm s}^2 -  \taus^2\right){\hat{u}\gs \hat{v}^*\gs + \hat{u}\gs^* \hat{v}\gs \over 2}\Big] \label{eq:epiuv}
\end{subeqnarray} 
where 
$d(\omega) =  [1+\nu_{\rm s}^2 + \taus^2]^2 - 4 (\nu_{\rm s} \taus)^2$
In eqns.\ (\ref{eq:epiamp}a--b), terms in the numerator $\propto \imath \nu_{\rm s}( \hat{u}\gs \hat{v}^*\gs - \hat{u}\gs^* \hat{v}\gs)$ are dropped since they are odd in $\nu_{\rm s}$ and will vanish on integration over frequency.

Since the amplitudes have the form, \eg $|\hat{u}\ps|^2 = a |\hat{u}\gs|^2 + b  |\hat{v}\gs|^2 + c (\hat{u}\gs \hat{v}^*\gs + \hat{u}\gs^* \hat{v}\gs)/ 2$, we generalize the particle power spectrum from \Eq{eq:powerp} to:
\begin{equation} 
\hat{E}_{\mathrm{p},x}(\omega) = \left[a \brak{\ug^2} + b \brak{\vg^2} + c \brak{\ug \vg}\right]\hat{P}(\omega)
\end{equation} 
and similarly for the $y$ power and $x$--$y$ correlations.  The particle energies are derived as before, \eg $\brak{\up^2} = \int_{-\infty}^\infty \hat{E}_{\mathrm{p},x} d\omega$, to give:
\begin{subeqnarray}\label{eq:episq}
\brak{\up^2} &=& {1\over d_2}\Big\{\left[1+ \St(1+\taus^2/2) \right]\brak{\ug^2} +\nonumber\\
&& 2\tau_s^2(2 + \St)\brak{\vg^2}  + 2\tau_s(2 + \St)\brak{\ug \vg}\Big\} \\
\brak{\vp^2} &=&{1\over d_2} \Big\{ {\tau_s^2(2 + \St) \over 8}\brak{\ug^2} + \left[1 + \St(1+\taus^2/2) \right]\brak{\vg^2} - \nonumber \\
&& \tau_s(2 + \St)\brak{\ug \vg}/2 \Big\} \\
\brak{\up \vp} &=&{1\over d_2} \Big\{-{ \tau_s(2 + \St) \over 4}\brak{\ug^2} +  \tau_s(2 + \St) \brak{\vg^2}  + \nonumber\\
&&\left(1+ \St -\taus^2\right)\brak{\ug \vg}  \Big\}
\end{subeqnarray} 
where $d_2 \equiv (1+\taus^2)\left[(1+\St)^2 + \taus^2\right]$.

Since the expressions are fairly cumbersome it is fruitful to investigate limiting cases.  \Tab{tab:ordering} lists the limiting forms of $\delta\up/\delta\ug \equiv (\brak{\up^2}/\brak{\ug^2})^{1/2}$, using the turbulence model in \Eq{eq:turbmodel} for the other components of gas forcing.   For more general analysis, we first consider  $\taus \ll 1$, i.e. the left half of \Fig{fig:tets}.  The equations simplify considerably to:
\begin{equation} \label{eq:epitight}
{\brak{\up^2}\over \brak{\ug^2}} = {\brak{\vp^2} \over \brak{\vg^2}} = {\brak{\up\vp} \over \brak{\ug\vg}} = {1 \over 1+\St}\, .
\end{equation}
There is no epicyclic motion since particles are tightly coupled to gas orbits, and the response scales directly with the forcing.\footnote{A possible exception occurs if, e.g.\ the azimuthal turbulent forcing is much stronger than the radial.}  The scaling with $\St$ agrees with \Eq{eq:upsq} in the absence of orbital dynamics.  

Next consider the limit of short eddy times, $\taue \ll1$ and $\St \gg 1$ in regions II and III, which we will compare to our Fokker-Planck results in \S\ref{s:FP}.  Here the particle response is,
\begin{subeqnarray} \label{eq:shorteddy}
\brak{\up^2} &=& {\left(1+{\taus^2 \over 2}\right)\brak{\ug^2}+  2\taus^2 \brak{\vg^2} + 2\taus \brak{\ug\vg} \over \St(1+\taus^2)} \nonumber\\
&=& {\brak{\ug^2}\over \St} {1+5\taus^2/2 \over 1 + \tau_s^2} \\
\brak{\vp^2} &=& { {\taus^2\over 8} \brak{\ug^2} + \left(1+{\taus^2 \over 2}\right)\brak{\vg^2}-  \taus \brak{\ug\vg}/2  \over \St(1+\taus^2)}\nonumber\\
 &=& {\brak{\ug^2}\over \St} {1+5\taus^2/8 \over 1 + \tau_s^2} \\
\brak{\up \vp} &=& { -{\taus \over 4} \brak{\ug^2}+\taus \brak{\vg^2} +
\brak{\ug\vg} \over \St(1+\taus^2)}
\nonumber\\ 
&=&
  {\brak{\ug^2}\over \St} {3\taus/4 \over 1 + \tau_s^2} 
\end{subeqnarray}
where the final equalities assume isotropic turbulence with $\brak{\ug^2} = \brak{\vg^2}$ and ignore  $ \brak{\ug \vg}$, consistent with \Eq{eq:turbmodel}.\footnote{
We have dropped a term $-\tau_{\rm s}\tau_{\rm e}\brak{\ug\vg}$ from the numerator
of the middle expression in equation (\ref{eq:shorteddy}c) because it is always smaller than the other
terms in the numerator for $\tau_{\rm e}\ll 1$.
}
  For small $\taus$ equations (\ref{eq:shorteddy}) recover the $\St \gg 1$ limit of \Eq{eq:epitight}.  For $\taus \gg 1$ we recover epicyclic motion.  The radial RMS speeds are twice the azimuthal RMS speeds, whether the forcing is radial, azimuthal or combined.\footnote{Random velocities are measured relative to the local circular speed.  Relative to a fixed guiding center the elongation of epicycles is opposite; azimuthal speeds are twice radial \citep{bt}.}  Equation (\ref{eq:shorteddy}c) shows that radial turbulent forcing produces a negative $\brak{\up \vp}$, i.e.\ inward transport of particle angular momentum. Azimuthal forcing has the opposite sign, and dominates for our assumption of isotropy.  

Region IV exhibits epicyclic oscillations since $\varOmega^{-1}$ is the most rapid timescale.  The ratio of RMS particle  to turbulent speeds scale at $\sqrt{\St}/\taus = 1/\sqrt{\taus \tau_{\rm e}}$ in agreement with the physical arguments in \S\ref{s:bigtaue}.  Region V is rather complicated since even though $\taus > 1$ particles do not complete epicycles as they are tightly coupled to eddies, $\St < 1$.

\begin{figure}[htbp] 
\hspace{-1cm}  \includegraphics[width=4in]{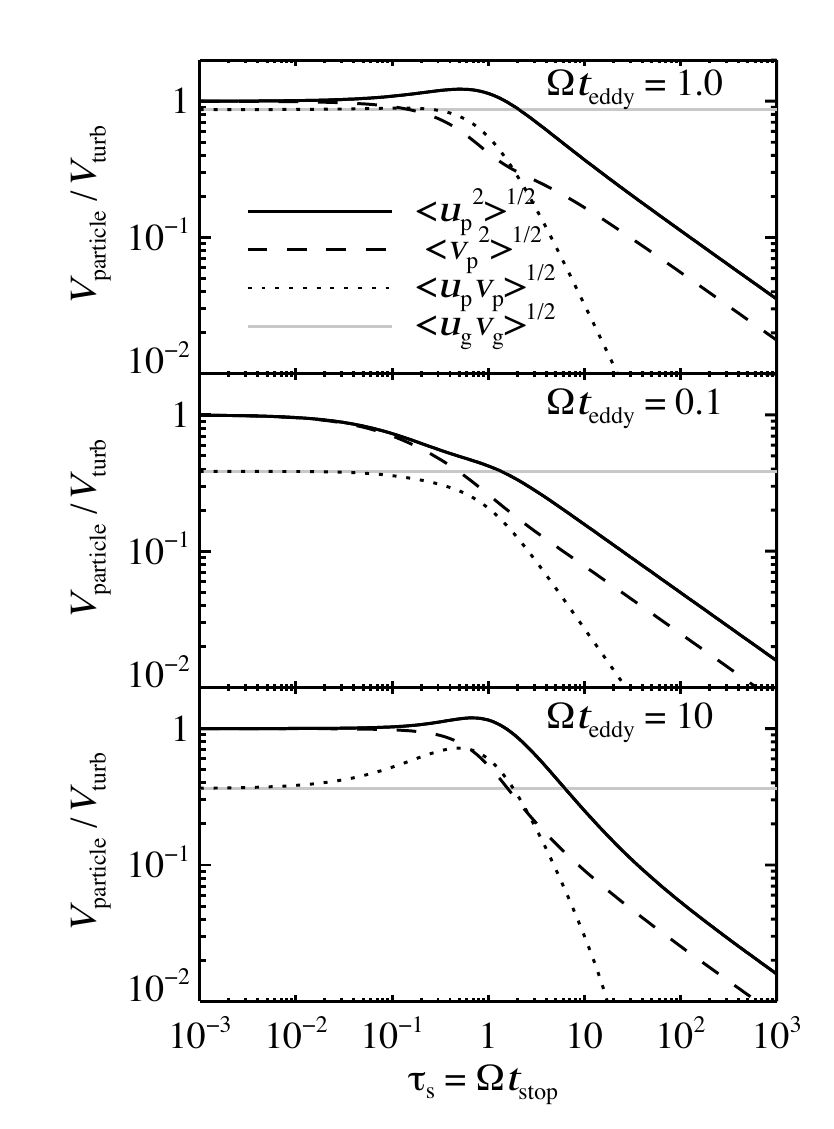} 
   \caption{Particle RMS speeds (radial, \emph{solid curves}; azimuthal, \emph{dashed curves}), relative to the turbulent RMS speed, $V_{\rm turb} = \sqrt{\brak{\ug^2 + \vg^2}/2 }$, versus stopping time for three different values of the eddy time ($\te = 1.0/\varOmega$,  $0.1/\varOmega$, and $10.0/\varOmega$ in the \emph{top, middle}, and \emph{bottom} panels).  The (square root of) the particle velocity correlations is plotted (\emph{dotted curves}).  The turbulent gas correlations (\emph{grey lines}) transport angular momentum, but have a small effect on the particle velocities (and only near $\taus = 1$).  See \S\ref{s:alpha} for discussion.}
   \label{fig:upvp}
\end{figure}

\subsubsection{Numerical Evaluation and Visual Interpretation of Velocities}\label{s:alpha}

\Fig{fig:upvp} plots RMS velocities and velocity correlations from the (square root of) equations (\ref{eq:episq}) for the turbulence model of \Eq{eq:turbmodel}.  This model assumes that velocities are isotropic, $\brak{u\gs^2} = \brak{v\gs^2}$, and estimates a maximum correlation, $\brak{u\gs v\gs}$ (grey curves).  Speeds are normalized to the characteristic turbulent speed, $V_{\rm turb} \equiv (\brak{\ug^2 + \vg^2}/2)^{1/2}$, and results are shown for $\tau_{\rm e} = 0.1,1.0$, and $10.0$ (middle, top, and bottom panels respectively).

For $\taus \ll 1$, particle RMS speeds simply follow the gas forcing, here the isotropic value $V_{\rm turb}$.   For $\taus \gg 1$, the RMS speeds fall as $\taus^{-1/2}$, and the ratio of radial (solid curves) to azimuthal  (dashed curves) speeds approaches the epicyclic value of two.  As $\taue$ deviates from unity (becoming either larger or smaller) the RMS speeds at large $\taus$ drop roughly as $\sqrt{\taue/(1+\taue^2)}$.  Similar behavior was seen for vertical speeds in \Fig{fig:wpts}.  Thus the RMS velocities of large particle are a good indicator of whether the eddy time differs from unity.

The speeds near $\taus \approx 1$ are sensitive to the eddy time and the $\brak{u\gs v\gs}$ correlations.  For $\taue = 1.0$, the ``bump" in radial speed near $\taus = 1$ is due to the high value of $\brak{u\gs v\gs}$ (otherwise the radial velocities would be flat up to $\taus = 1$).  For $\taue = 0.1$ there is a noticeable drop in speeds by $\taus = 1$, because with $\St = \taus/\taue = 10$ particles have begun to decouple from eddies.   For $\taue = 10$ the radial velocity bump exists even without gas correlations, which are weak here.  The difference in speeds near $\taus \approx 1$ discriminates between large and small eddy times.   We emphasize that the turbulent transport of angular momentum by Reynolds stresses $\brak{\ug\vg}$ has only a small effect on particle speeds, and then only near $\taus \approx 1$.


\subsubsection{Radial Particle Diffusion}
Radial diffusion\footnote{We do not here consider azimuthal diffusion.  It is less interesting since orbital shear is much more effective at spreading non-axisymmetric structure.} of particles can be calculated as in \Eq{eq:Dp} to give
\begin{eqnarray} 
D_{\mathrm{p},x} &=& \int_0^\infty d \tau\int_{-\infty}^\infty d \omega \hat{E}_{\mathrm{p},x}e^{-\imath \omega t}=\pi \hat{E}_{\rm p}(0) \nonumber \\
&=& \te { \brak{\ug^2} + 4 \taus^2 \brak{\vg^2} + 4\taus  \brak{\ug \vg} \over (1+\taus^2)^2}  \label{eq:Dpx}
\end{eqnarray}
The $\taus \ll 1$ limit recovers the result of \Eq{eq:Dp} that particles of any size diffuse as a gas tracer \emph{when orbital motions are neglected}.   
Azimuthal forcing has a negligible effect since tightly coupled particles quickly adjust to the local orbital speed. For $\taus \gg 1$, radial diffusion is dominated by the azimuthal forcing, $D_{\mathrm{p},x} = 4\brak{\vg^2}\te/\taus^2$, since angular momentum perturbations are needed to change semimajor axes. 
The scaling for loose coupling agrees with the estimate of \Eq{eq:Dpest}.  Correlated gas fluctuations, $\brak{\ug\vg}$, are a minor correction to radial particle diffusion, and only near $\taus \sim 1$ (as with the velocities).  The limiting forms of \Eq{eq:Dpx} are listed in \Tab{tab:ordering} for the turbulence model in \Eq{eq:turbmodel}.

The radial Schmidt number can be expressed for homogenous isotropic turbulence $ \brak{\ug^2}= \brak{\vg^2}$, $  \brak{\ug\vg}=0$ as:
\begin{equation} \label{eq:Scx}
\Sc_x \equiv {D_{\mathrm{g},x} \over D_{\mathrm{p},x}} = {(1+\taus^2)^2 \over 1 + 4 \taus^2}\, ,
\end{equation}
where $ D_{\mathrm{g},x} = \brak{\ug^2}\te$.  This is consistent with, but provides order unity corrections to the rough estimate in \Eq{eq:diskest}.  It is interesting that this result is completely independent of the eddy time.

\begin{figure}[htbp] 
\hspace{-1cm}   \includegraphics[width=4in]{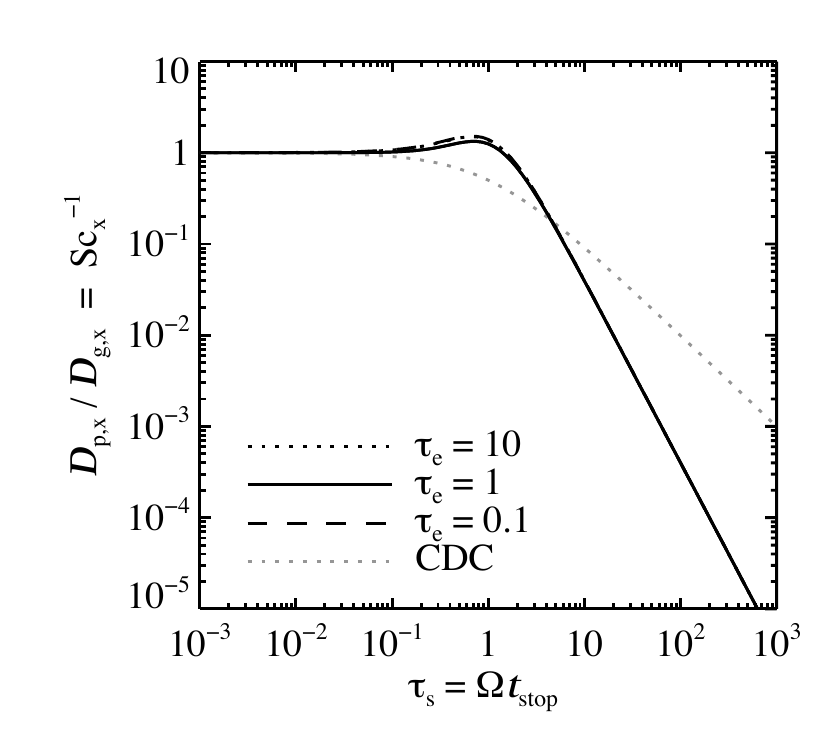} 
   \caption{Radial particle diffusion, relative to gas diffusion, versus stopping time.  The black curves [from \Eqs{eq:Dpx}{eq:turbmodel}] overlap despite different choices of eddy time.  For tight coupling ($\taus \ll 1$) particles diffuse as effectively as gas.  For loose coupling ($\tau_{\rm s} \gg 1$) particle diffusion falls as $\taus^{-2}$.     The CDC result, $\Sc = 1 + \St$ with $\tau_{\rm e} = 1$ (\emph{grey dotted curve}) overestimates the diffusion of large ($\taus \gg 1$) particles and slightly underestimates diffusion of marginally coupled solids.}
   \label{fig:Dp}
\end{figure}

\Fig{fig:Dp} plots the radial particle diffusion, $D_{\mathrm{p},x}$, relative to the diffusion of gas, $D_{\mathrm{g},x}$.  The black curves use \Eq{eq:Dpx} and the turbulence model in \Eq{eq:turbmodel}.  Since the correlations $\brak{u\gs v\gs}$ are a minor contribution, the black curves overlap with each other and with \Eq{eq:Scx} (or rather its inverse, not plotted).  The overlap only occurs because $\brak{u\gs^2} = \brak{v\gs^2}$ in all cases.  If the velocities were anisotropic then the diffusion for $\taus \gg 1$ would vary as $\brak{v\gs^2}/\brak{u\gs^2}$, i.e.\ $D_{\mathrm{p},x}$ would be larger (smaller) if azimuthal kinetic energy exceeds (falls below) the radial contribution.

The CDC value for $\Sc$, \Eq{eq:ScCDC}, is plotted for reference (dotted grey curve).  The physical shortcomings of this result (neglect of epicyclic oscillations and particle mean free path) are discussed in \S\ref{s:CDC}.  While the CDC value over-predicts diffusion for very large particles $\taus \gg 1$, it tends to under-predict diffusion for marginal coupling, $\taus \sim 1$.

\subsubsection{Monte Carlo Diffusion Simulation}\label{s:diffsim}
\begin{figure}[htbp] 
   \includegraphics[width=3.6in]{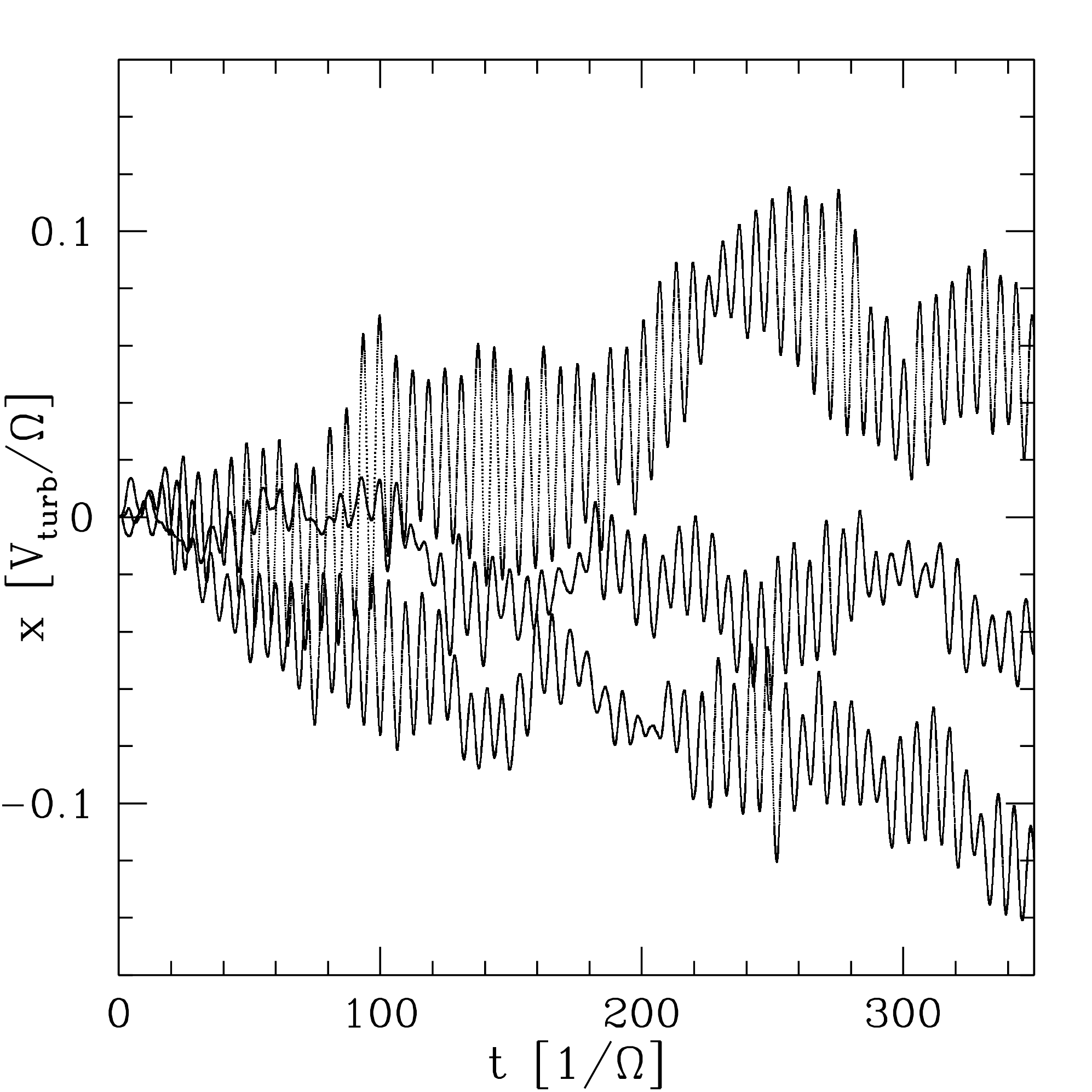} 
   \caption{A Monte Carlo simulation of the radial diffusion of three particles undergoing epicyclic motion with $\taus = 10^2$ and $\tau_{\rm e} = 10^{-2}$. The radial coordinate $x$ is normalized to the RMS turbulent gas speed divided by the Kepler frequency. See \S\ref{s:diffsim} for details.}
   \label{fig:diff}
\end{figure}

Testing our results with detailed turbulence simulations (or experiments) is beyond the scope of this paper.  Instead we visualize the combination of diffusion and epicyclic motion with the Monte Carlo simulation shown in \Fig{fig:diff}.  We integrated equations (\ref{eq:Hills})
numerically, with $\tau_s=10^2$ and $\tau_e=10^{-2}$ (region III of Figure \ref{fig:tets}).  As a crude model
for the forcing by the gas, we set $u\gs(t)=\pm 1$ and $v\gs(t)=\pm 1$ for a time approximately $t_{\rm eddy}$.
More precisely, we updated $u\gs$ and $v\gs$ at every time step with probability $dt/t_{\rm eddy}$, where
$dt$ is the time step, and the decision to update was implemented with a random number generator.  If
an update was warranted, the sign of $u\gs$ and of  $v\gs$ was each randomly selected.
The three curves in the figure show three realizations of the random number generator.
For the oscillation amplitudes, equation (\ref{eq:shorteddy}a) 
predicts that $\langle u\ps^2\rangle=(5/2)\langle u\gs^2\rangle\tau_e/\tau_s=(0.016)^2$, implying
that the peak-to-peak amplitude of $x$ should be $0.016\sqrt{8} \varOmega^{-1}=0.04\varOmega^{-1}$,
in agreement with the oscillations seen in the figure.  For the diffusion, equation (\ref{eq:Dpx}) 
predicts the diffusion constant $D_{\mathrm{p},x}=t_{\rm eddy}(4/\tau_s^2)\langle v\gs^2\rangle=\varOmega^{-1}(0.002)^2$, implying that $x\simeq ({2D_{\mathrm{p},x}t})^{1/2}=\varOmega^{-1}0.05$ at $t=300\varOmega^{-1}$,
in agreement with the diffusion seen in the figure.

\section{Fokker-Planck Equation}\label{s:FP}

When the eddy correlation time $t_{\rm eddy}$ is the smallest timescale, 
the random gas motions give short
uncorrelated kicks to the particles which can be described by 
a Fokker-Planck equation.  The Fokker-Planck equation has been used extensively to study Brownian motion \citep{cha43}.  While the uncorrelated collisions in Brownian motion lead to thermal equipartition, the correlated motion of turbulent eddies excite particles to much higher speeds.  CFP use the Fokker-Planck technique to study vertical settling.  By solving the in-plane Fokker-Planck equation we reproduce the results of
  \S \ref{s:Hepi} in the small $t_{\rm eddy}$ limit, adding
  confidence in our results.
  
  It greatly simplifies the analysis to switch to indicial notation, re-writing the equations of motion for a particle given in \Eq{eq:Hills} as
\begin{subeqnarray}
{d x_i\over dt} = u_i-{3\over 2}\varOmega x_1 \delta_{iy} 
\label{eq:dxdt}
\\
{du_i\over dt}=A_{ij}u_j+{u_{g,i}\over t_{\rm stop}}
\label{eq:dudt}
\end{subeqnarray}
 where $i,j\in \{x,y\}$,
 summation over repeated indices is implied, and
 \begin{equation}
 A_{ij}\equiv
\left(\begin{array}{cc} -t_{\rm stop}^{-1}& 2\varOmega \\ -\varOmega/2 & -t_{\rm stop}^{-1}\end{array}\right) \ .
 \end{equation}
We suppress the subscript p from $x_i,u_i$.

The Fokker-Planck equation for the particle distribution function $f(x_i,u_i,t)$ reads
\begin{eqnarray}
\partial_t f = 
-\partial_i \left( {\left\langle \Delta x_i \right\rangle \over \Delta t} f \right)
&-&\partial_{u_i}\left({ \left\langle \Delta u_i\right\rangle\over \Delta t}f\right)\nonumber\\
&+&{1\over 2} \partial^2_{u_iu_j}
\left({\left\langle{\Delta u_i\Delta u_j}
\right\rangle \over \Delta t}f\right)
\end{eqnarray}
(e.g., equation 8-53 in \citealp{bt}).  The angled brackets
denote an ensemble average over the stochastic variable $u_{g,i}$, and
$\Delta x_i,\Delta u_i$ denote the change in $x_i,u_i$ over the time interval $\Delta t$,
in the limit that $\Delta t\rightarrow 0$. (To be more precise, $\Delta t$ must be larger than $t_{\rm eddy}$, 
but much smaller than all other timescales).
We have dropped the term containing $\langle \Delta x_i\Delta x_j \rangle /\Delta t $ because, 
from equation (\ref{eq:dxdt}a), it is proportional to $\Delta t$, and hence vanishes for $\Delta t\rightarrow 0$.
To calculate  $\langle \Delta u_i\Delta u_j \rangle /\Delta t $, we 
may drop the $A_{ij}u_j$ on the right-hand side of equation (\ref{eq:dudt}b)
because it only contributes a term of order $\Delta t$ to  $\langle \Delta u_i\Delta u_j \rangle /\Delta t $;
we are left with
\begin{equation}
\Delta u_i = {1\over t_{\rm stop}}\int_t^{t+\Delta t} u_{g,i}(t')dt' \ ,
\end{equation}
implying
\begin{eqnarray}
{\left\langle
\Delta u_i\Delta u_j
\right\rangle
\over 
\Delta t}
&=&
{1\over \Delta t}
{2\over t_{\rm stop}^2}
\int_t^{t+\Delta t}\int_t^{t'}
\left\langle
u_{g,i}(t')u_{g,j}(t'')
\right\rangle
dt''dt'
 \nonumber \\
&=&
{2t_{\rm eddy}\over t_{\rm stop}^2}
\langle
u_{g,i}u_{g,j}
\rangle \, ,
\end{eqnarray}
in agreement with eq.\ (9) of CFP.  The remaining averages are given by equations (\ref{eq:dxdt}) as
\begin{subeqnarray}
{
\left\langle
\Delta x_i
\right\rangle
\over \Delta t
}
&=& \langle
u_i
\rangle
-{3\over 2}\Omega\langle x_1\rangle\delta_{iy}
\\
{
\left\langle
\Delta u_i
\right\rangle
\over \Delta t
}
&=&
A_{ij}\langle u_j\rangle
\end{subeqnarray}
The Fokker-Planck equation then becomes
\begin{eqnarray}
{\tilde{\partial}_t}f+u_i\partial_i f 
=-\partial_{u_i}(A_{ij}u_jf)+D_{u_iu_j}\partial_{u_iu_j}^2f \ ,
\label{eq:fok}
\end{eqnarray}
 suppressing angled brackets and defining
\begin{equation}\label{eq:veldiff}
D_{u_iu_j}\equiv 
{t_{\rm eddy}\over t_{\rm stop}^2}\langle u_{g,i}u_{g,j}\rangle
\end{equation}
as the velocity diffusion, and
 $\tilde{\partial}_t \equiv \partial_t - (3/2)\varOmega x\partial_y$ which reduces to a simple time derivative for an axisymmetric ($\partial_y f=0$) distribution function.

We solve the Fokker-Planck equation (\ref{eq:fok}) by taking velocity moments, defining
  the mass per unit area, mean velocity, and peculiar velocity tensors:
\begin{subeqnarray}
\varSigma &\equiv& \int f   du_x du_y \label{eq:Sigma} \\
U_i &\equiv& \varSigma^{-1} \int u_if du_xdu_y \\
\Pi_{ij}&\equiv& \varSigma^{-1}\int 
(u_i-U_i)(u_j-U_j) f du_xdu_y \\
\Xi_{ijk}&\equiv&\varSigma^{-1} \int 
(u_i-U_i)(u_j-U_j)(u_k-U_k) {f du_xdu_y}\qquad \  \end{subeqnarray}
Multiplying equation (\ref{eq:fok}) by $\{1,u_i,u_iu_j\}$, and then integrating
out $u_x,u_y$, and $y$ yields
\begin{subeqnarray}
\tilde{\partial}_t \varSigma+
\partial_i\varSigma U_i &=& 0 \, ,
\label{eq:rt}
\\
\tilde{\partial}_tU_i
+U_j\partial_jU_i&=&-\varSigma^{-1}\partial_j(\varSigma\Pi_{ij})
+A_{ij}U_j\, ,
\label{eq:ut}
 \\
\tilde{\partial}_t\Pi_{ij}+U_k\partial_k\Pi_{ij} &+&
\Pi_{ik}\partial_kU_j+\Pi_{jk}\partial_kU_i
 + \nonumber \\
\varSigma^{-1}\partial_k(\varSigma\Xi_{kij})
&=&A_{ik}\Pi_{kj}+A_{jk}\Pi_{ki}+2D_{u_iu_j}\, .
\label{eq:st}
\end{subeqnarray}

We henceforth consider the axisymmetric case ($\partial_y = 0$) for simplicity.  As shown  below,
the left-hand sides of equations (\ref{eq:ut}) and (\ref{eq:st}) are typically negligible.
Assuming for now that this is true,   equation (\ref{eq:st}) can be solved
for the velocity dispersion,
\begin{eqnarray}
\left(\begin{array}{c}\Pi_{xx} \\ \Pi_{xy} \\ \Pi_{yy}\end{array}\right)
=
t_{\rm stop}
\left(\begin{array}{ccc}1 & -2\tau_s & 0 \\\tau_s/4 & 1 & -\tau_s \\0 & \tau_s/2 & 1\end{array}\right)^{-1}
\left(\begin{array}{c}D_{u_xu_x} \\D_{u_xu_y} \\D_{u_yu_y}\end{array}\right)&&\quad
 \nonumber \\
=
{t_{\rm stop}\over 1+\tau_s^2}
\left(\begin{array}{ccc}1+\tau_s^2/2 & 2\tau_s & 2\tau_s^2 \\ -\tau_s/4 & 1 & \tau_s \\ \tau_s^2/8 & -\tau_s/2 & 1+\tau_s^2/2\end{array}\right)
\left(\begin{array}{c}D_{u_xu_x} \\D_{u_xu_y} \\D_{u_yu_y}\end{array}\right) && \quad
\label{eq:sigmaij}
\end{eqnarray}
which recovers 
equations (\ref{eq:shorteddy}) once we identify $\{\Pi_{xx},\Pi_{xy},\Pi_{yy}\} = \{\brak{\up^2},\brak{\up\vp},\brak{\vp^2}\}$ and apply \Eq{eq:veldiff}.

Equation (\ref{eq:ut})  can be solved for the mean velocity
\begin{eqnarray}
\left(\begin{array}{c}U_x \\ U_y\end{array}\right)=
-{1\over \varSigma}
{ t_{\rm stop} \over 1 + \tau_s^2}\left(\begin{array}{cc}1 & 2\tau_s \\-\tau_s/2 & 1\end{array}\right)\partial_x
\left(\begin{array}{c}\varSigma \Pi_{xx} \\ \varSigma \Pi_{xy}\end{array}\right)\, .
\label{eq:ubar}
\end{eqnarray}
We now assume that $\varSigma$ varies on a shorter length scale than $\ts$ or $\Pi_{ij}$ do.  The diffusion equation, from equation (\ref{eq:rt}), takes the simple form
\begin{equation}
\partial_t\varSigma=D_{\mathrm{p},x}\partial_x^2\varSigma
\end{equation}
where 
\begin{eqnarray}
&&D_{\mathrm{p},x}=
{t_{\rm stop}\over 1+\tau_s^2}\left(\Pi_{xx}+2\tau_s\Pi_{xy}\right)  \nonumber \\
&&\quad =
\left(
{t_{\rm stop}\over 1+\tau_s^2}
\right)^2\left(
D_{u_xu_x}+4\tau_sD_{u_xu_y}+4\tau_s^2D_{u_yu_y}
\right) \quad \label{eq:DpxFP}
\end{eqnarray}
in exact agreement with \Eq{eq:Dpx}.

We now determine the conditions under which  the left-hand sides of equations (\ref{eq:ut}b,c) 
may be neglected.  Let $L$ be the length scale (in the $x$-direction) over which ln$\varSigma$ varies, and 
let $T$ be the timescale.  
Then $U_i$ also varies on length- and time-scales $L$ and $T$ (eq.\ [\ref{eq:ubar}]), 
with amplitude $U_i\sim L/T$ (eq.\ [\ref{eq:rt}]), and
$\Pi_{ij}$ is constant (eq.\ [\ref{eq:sigmaij}]), with $\Pi_{ij}\sim C^2$, where $C$ is the typical random speed.  The left-hand side of equation (\ref{eq:ut}) is smaller than the first term on the right-hand side when $L/T\ll C$.  As long as this inequality holds, the two terms on the right-hand side tend to equalize, with $C^2T/L^2\sim A_{ij}$, and the terms
on the left-hand side of equation (\ref{eq:st}) (other than the $\Xi_{kij}$ term) are smaller than 
the ``$A\Pi$'' terms on the right-hand side by $(L/CT)^2\ll 1$; the $\Xi_{kij}$ is smaller than 
the ``$A\Pi$'' terms by $L/CT\ll 1$.  In sum, our approximations are valid as long as $L/T\ll C$, i.e.,
that the mean speed of the particles is less than their random speed.  As long as this holds, the timescales
for both the mean speed and the random speed to reach their quasi-steady-state values 
of equations (\ref{eq:sigmaij}) and (\ref{eq:ubar})
are much 
shorter than the diffusion timescale $T$.

We return briefly to the discussion in \S\ref{s:vertapp} of whether particle concentration relative to gas is the fundamental quantity diffused by turbulence.  Equation (\ref{eq:ubar}) shows that, for $\taus \ll 1$ (to ignore the complexities of orbital motions), $\varSigma \Pi_{xx}$ is the diffused quantity, which is not proportional to relative concentration.  However this result is restricted to $\ts \gg \te$.  The standard assumption that diffusion levels concentration gradients holds for small particles tightly coupled to eddies.  Since this limit is not treated by the Fokker-Planck approach, there is no inconsistency.

\section{Neglected Effects}\label{s:neglect}
\subsection{Collisions}\label{s:coll}
Interparticle collisions are neglected both to simplify the analysis and because they are often a small correction to the dynamics, when compared to gas drag.  The collision time is
\begin{equation} 
t_{\rm coll} = {2 \rho_\bullet a \over 3 \rho\ps v_{\rm rel}}
\end{equation}
where $v_{\rm rel}$ is the relative velocity between particles and $\rho_\bullet \sim 1$ -- 3 g/cm$^3$ is the internal solid density.  For collisions between two species $i$ and $j$,\footnote{Our stirring model assume a single particle size.   Since we work in the test particle limit with no particle collisions, our results can be applied to a dispersion of particle sizes with no additional approximation.} the collision time for species $i$ due to $j$ is set by the space density of $j$ and particle radius, $a  \rightarrow 2 a_j^3/(a_i^2 + a_j^2)$.

If we compare to the stopping time in the Epstein regime, $\ts^{\rm Ep} = \rho_\bullet a /( \rho\gs c\gs)$, we see that
\begin{equation} 
{\ts^{\rm Ep} \over t_{\rm coll}} \simeq {\rho_{\rm p} \over \rho_{\rm g}}{v_{\rm rel} \over c\gs} \ll 1
\end{equation} 
where inequality follows since both fractions are less than unity in a smooth turbulent disk.\footnote{If significant clumping to $\rho\ps > \rho\gs c\gs/v_{\rm rel} >> \rho\gs$ occurs, from e.g.\ streaming and/or gravitational instabilities, then collisions would become dynamically significant.}  The particle density should be less than the gas density for our test particle approach (see \S\ref{s:feedback}) and the relative motions between particles (whether induced by turbulence or aerodynamic drift) are always subsonic.

In the Stokes regime (when the particle radius exceeds the gas mean free path, $\lambda$, but the Reynolds number of the flow around the particle remains below unity) collisions become relatively more important as the stopping time increases by $4a/(9\lambda) > 1$.  However turbulent drag sets in before collisions dominate in gas rich disks.  For turbulent drag, $\ts^{\rm turb} \approx \rho_\bullet a/(\rho\gs \Delta V)$, where $\Delta V$ is the relative velocity of particles and gas.  By this point (Appendix \ref{s:sizes}) particles are sufficiently decoupled from the gas that $v_{\rm rel} \approx \Delta V$, and
\begin{equation} 
{\ts^{\rm turb} \over t_{\rm coll}} \approx {\rho_{\rm p} \over \rho_{\rm g}} \, 
\end{equation} 
gives the result (for turbulent drag) that drag is more significant than collisions if the local mass in gas exceeds that in particles.  We showed above that this criterion is over-restrictive in the linear drag regimes (Stokes and Epstein).

\subsection{Spatial Spectrum of Eddies and Crossing Trajectories}\label{s:CT}
Unlike the detailed treatments pioneered by VJMR, we neglect the spatial spectrum of turbulence.  This is valid if the time to cross an eddy, $t_{\rm cross} = \ell_{\rm eddy} / \Delta V$ (with $\ell_{\rm eddy}$ the largest eddy scale and $\Delta V$ the relative motion of particles and gas), is small compared to either the eddy time (so that a particle never sees the spatial extent of an eddy) or the stopping time (so that a particle is fully accelerated before crossing an eddy).  Thus spatial homogeneity is a good approximation for
\begin{equation} \label{eq:crossing}
t_{\rm cross} > \rm{Min}(\ts,\te)\, .
\end{equation} 
If $\Delta V$ arises from turbulent forcing, then this criterion is satisfied.   For tight coupling, $\St \ll 1$, tight coupling to eddies gives $\Delta V \ll \delta V\gs$.  Thus $t_{\rm cross} /\te = \delta V\gs/\Delta V \gg 1$, and  $t_{\rm cross} /\ts$ is even larger, so \Eq{eq:crossing} is readily satisfied.  For loose coupling, $\St \gg 1$, the particle RMS motions are small so $\Delta V \sim \delta V\gs$, the gas forcing.  Thus $t_{\rm cross}/\te \sim \ell_{\rm eddy}/(\delta V\gs\, \te) = 1$, and \Eq{eq:crossing} is only marginally satisfied.

Orbital oscillations further help to confine particles to an eddy because the size of  epicyclic or vertical oscillations, $\ell_{\rm epi}$, is smaller than the eddy scale:
\begin{equation} 
{\ell_{\rm epi} \over \ell_{\rm eddy}} = {\delta V\ps \over \varOmega \ell_{\rm eddy}} = {\delta V\ps \over \delta V\gs} \sim {1 \over \sqrt{\St}} \gg 1\, .
\end{equation} 
The above assumes $\tau_{\rm e} \sim 1$ and $\St \gg 1$ to that particle complete epicycles.  We conclude that spatial homogeneity is a safe approximation for pure turbulent forcing, with (as usual) possible corrections near $\taus = 1$.
 
However if $\Delta V$ is dominated by net motion of particles relative to gas eddies, the spatial extent of the eddies can be significant.  CDC follow \citet{csa63} in describing this as the ``crossing trajectories effect" (hereafter CTE).   Azimuthal and radial drift of particles due to radial pressure gradients, $\p P/\p r$, is the most likely source of net motion.\footnote{Vertical settling can be neglected as it is smaller than particle (and thus gas) RMS speeds from \Eq{eq:settRMS}.}  The 
radial and azimuthal drift speeds in the test particle limit are \citep{nsh86}:
\begin{eqnarray} 
\Delta V_r &=& -{2\taus \over 1+\taus^2} \eta v_{\rm K}\, ,  \label{eq:vr} \\
\Delta V_\phi &=& {\taus^2 \over 1+\taus^2} \eta v_{\rm K}\, ,\label{eq:vphi}
\end{eqnarray} 
where the dimensionless pressure support parameter,
\begin{equation} \label{eq:eta}
\eta \equiv -{\p P/\p \ln r \over 2 \rho\gs v_{\rm K}^2} \approx 1.5 \left({c\gs \over v_{\rm K}}\right)^2\, ,
\end{equation}  
and $v_{\rm K}$ is the Keplerian orbital speed.

For loose coupling, azimuthal drift at the full $\Delta V_\phi \approx \eta v_{\rm K}$ dominates the net motion.
Eq.\ (\ref{eq:crossing}) will be satisfied (and the CTE will be ignorable) if turbulence is sufficiently strong so that $\delta V\gs > \eta v_{\rm K}$.  If we model the turbulence as $\delta V\gs \sim \sqrt{\alpha}c\gs$, then $\alpha > \eta \sim 10^{-3}$ is required, i.e.\ reasonably, but not exceptionally strong turbulence.  For tight coupling, radial drift dominates relative motion with $\Delta V_r \approx \eta v_{\rm K} \taus \ll \eta v_{\rm K}$.  Eq.\ (\ref{eq:crossing}) is satisfied for $\delta V\gs > \eta v_{\rm K} \taus^2$ or $\alpha > \eta \taus^4$.  Thus even very weak turbulence ($\alpha \gtrsim 10^{-16}$ for mm-sized pebbles at 1 AU) will negate the CTE for $\taus \ll 1$.

When the CTE is relevant, our results should apply approximately for an effective eddy time, $\te'$, that is the smaller of the actual eddy time  and the crossing time,
\begin{equation} \label{eq:CTE}
 \te' \approx (1/\te^2+ 1/t_{\rm cross}^2)^{-1/2} \approx \te/\sqrt{1 + (\Delta V/\delta V_{\rm g})^2}\, .
 \end{equation} 
 This correction factor is similar to that used by CDC in their equations (43) and (B12).\footnote{Though CDC apply the correction for all $\St$, even if $\ts \ll \te'$, and there should be no effect if the particles are fully accelerated before crossing an eddy.}  Our prediction also appears consistent with VJMR, in particular Fig. 1 of the related \citet{vjmr78} (where our relative drift $\Delta V$ is their $V_{\rm L}$).

\subsection{Particle Feedback and Concentration}\label{s:feedback}
We model the passive response of particles to imposed turbulence, ignoring the aerodynamic feedback of particles on turbulent gas dynamics.  This restricts our analysis to local particle densities below the gas density, $\rho\ps \ll \rho\gs$.  This approximation is roughly valid for turbulence strong enough and/or particles small enough to prevent settling to such high densities.  Specifically, $\rho\ps/\rho\gs \simeq (\varSigma\ps/\varSigma\gs)(H\gs/H\ps) \lesssim 1$ requires
\begin{equation} \label{eq:feed}
\alpha_z \gtrsim \left(\varSigma\ps/\varSigma\gs\right)^2\taus \sim 10^{-4} \taus\, , 
\end{equation} 
from  \Eq{eq:Hpstd}, and a nominal $\varSigma\ps/\varSigma\gs \sim 10^{-2}$.  Enhancements of solid column densities relative to gas requires even stronger turbulence to remain in the test particle limit.  It is difficult to keep large particles from settling, but if only a small mass fraction resides in large particles, feedback effects might remain small.

Drag feedback effects are known to trigger turbulence by themselves.  Shear-induced instabilities from particle settling, i.e. Kelvin-Helmholtz instabilities, have been identified as a foil to the gravitational collapse of solids  (\citealp{stu80}, CDC). Recently it has been shown that the in-plane drift of solids and gas triggers instabilities and turbulence when feedback is included \citep{yg05,yj07}.  These ``streaming" instabilities are particularly relevant since they generate significant particle clumping \citep{jy07} even when shear is also included \citep{jhk06}.

In addition to the streaming instability, other effects cause particles to concentrate in turbulent flows, as discussed in the introduction.  Most concentration mechanisms are size dependent and transient.  (Appendix \ref{s:turbmodel} mentions the likely destruction long-lived vortices in 3-D).  Nevertheless non-diffusive behavior, especially on the integral scale of turbulence as seen in \citet{jhk06}, may cause discrepancies with our simplified turbulence model. 
In the absence of gravitational collapse diffusion should still occurs on long length and time scales.  Comparing our results to simulations (and parameter regimes) with different degrees of particle concentration will deepen our understanding of non-diffusive and other intermittent turbulent behaviors.  For instance treating the global pressure gradients (and hence radial drift) as a free parameter should modulate particle concentration.  Drift that is either too strong (blasting through any eddy) or too slow (reducing the pileups in an eddy time) should reduce particle concentration.

\section{Discussion}\label{s:Disc}
We study the diffusion of particles, and the excitation of their random velocities, by turbulent forcing in Keplerian gas disks.  We obtain an improved result for radial particle diffusion in \Eqs{eq:Dpx}{eq:DpxFP}, (identical formulae from independent techniques).  We express radial particle diffusion in an approximate form for isotropic turbulence in \Eq{eq:Scx} and even more crudely (good for quick estimates) in \Eq{eq:diskest},
\begin{equation}
\Sc_x \equiv {D_{\mathrm{g},x} \over D_{\mathrm{p},x}} = {(1+\taus^2)^2 \over 1 + 4 \taus^2} \sim 1 + \taus^2\, .
\end{equation}
Our result improves on the estimate of particle diffusion in CDC, $\Sc_{\rm CDC} = 1 + \St = 1 + \taus/\taue$, by including orbital dynamics and calculating diffusion for longer than a single eddy time (see \S\ref{s:CDC}).  The discrepancy with the CDC value is only significant for large particles with stopping times longer than the eddy or orbital time.  

Our new results for in-plane velocity dispersions, \Eq{eq:episq}, combine turbulent forcing  and epicyclic motions. Our results for vertical stirring in \S\ref{s:Hvert} agree with previous authors, including CDC (see their eq.\ [53]) and \citet{dms95} for tight particle coupling, and CFP for loose particle coupling.   We add a correction for arbitrary eddy times which reduces particle velocities (eq.\ [\ref{eq:wpsq}]) and scale heights (eq.\ [\ref{eq:Hpsq}]) for long eddy times.  We include simple, physically motivated prescriptions to  generalize the scale height result to include thorough mixing of small particles with gas (eq.\ [\ref{eq:Hpstrong}]) and to describe the reduction of the effective (Lagrangian) eddy time due to particle drift (eq.\ [\ref{eq:CTE}]).

We do not imagine that our results will significantly influence collisional velocities since the RMS velocities we calculate have a similar scaling to previous results.  Order unity corrections are possible for collisions involving marginally coupled particles, as orbital dynamics becomes important and the drift through eddies is maximized, but that calculation was beyond the scope of this work.

The rates of gravitational collapse of large particles, which are limited by mass diffusion when Toomre's $Q > 1$, will be increased.   Preliminary investigations indicate that the rate of dissipative gravitational collapse remains constant as $\taus \rightarrow \infty$ while the CDC result would predict a growth rate that falls as $\taus^{-1}$ (Youdin, in prep).  Other instabilities limited by particle diffusion may be influenced as well.

The detailed predictions of random velocities should prove useful in the interpretation of numerical simulations which couple particle dynamics to (usually MHD) turbulence.  CFP already confirmed their results for vertical oscillations with simulations of MRI driven turbulence.  It remains to perform numerical tests of in-plane motions, but note that CFP have already recovered the expected epicyclic velocity dispersions (radial speeds twice azimuthal) for $\taus = 10$ particles.  Ideally simulations of forced turbulence could investigate the effects of variable eddy times.

We can apply our results the the chemical gradients in the solar system.  A notable example is the spectroscopic zonation of the asteroid belt, but the issue is more general \citep{tay01}.  We consider, very roughly, the minimum size a particle must have
to drift radially faster than it diffuses, which is a crude measure of whether particles of 
a given size  could maintain zonation over a drift time.  Using the $\alpha$-parameterization  for turbulent gas diffusion, $D_{\mathrm{g},x} \equiv \alpha_x c\gs^2/\varOmega$, equations (\ref{eq:diskest}) and (\ref{eq:vr}) give the ratio of drift to diffusion time (over a scale of the disk radius $r$) as $t_{\rm drift}/t_{\rm diff} \sim \alpha_x/\tau_s$.  Any particle with $\taus > \alpha_x$ will diffuse slower than it migrates radially, i.e.\ a particle bigger than a centimeter for $\alpha_x \approx 0.01$ at $r \sim 2.5$ AU.  This simple result deserves two comments.  First it is the same criterion for a particle to be large enough to sediment to the midplane, from \Eq{eq:Hpstd} (as long as $\alpha_x \sim \alpha_z$ which holds at least to order-of-magnitude).  Second, since $\alpha_x \ll 1$ for subsonic turbulence, this fiducial particle size is tightly coupled, and could have been obtained equally well from the CDC result.  
Thus even though CDC overpredict the diffusion of large particles, the global implications are minor: large particles drift aerodynamically faster than they diffuse (with either result).

The reduced diffusion of heavy particles would slow the loss of particles from pressure maxima \citep{whi72}, including anti-cyclonic vortices, where radial drift is reduced.
Our improved result for diffusion is most relevant on smaller length scales, where diffusion rates are faster.  This is the case for the dissipative gravitational instabilities mentioned above.

\appendix
\section{Particle Sizes and Drag Regimes}\label{s:sizes}
To translate from $\taus$ to particle radius, $a$, we consider spherical particles of internal density $\rho_\bullet = 2 $  g/cm$^3$.  We use a standard minimum mass nebula model \citep{stu77b, hay81}, which has a gas mean free path (evaluated in the midplane) \citep{nsh86},
\begin{equation} \label{eq:mfp}
\lambda = \frac{\mu}{\rho_{\rm g} \sigma_{\rm mol}} \approx 1.2~ {\rm m}~ \varSigma_{150}^{-1} \left(\frac{r}{5 \rm AU}\right)^{2.75}\, ,
\end{equation} 
where $\mu = 3.9 \times 10^{-24}\,{\rm g}$ is the mean molecular weight, and
$\sigma_{\rm mol} = 2 \times 10^{-15}~ {\rm cm}^2$ is the molecular cross
section \citep{cc70}.  The gas surface density profile, $\varSigma\gs \propto r^{-3/2}$ is normalized to $\varSigma_{150} \equiv \varSigma\gs(5~\rm{AU})/(150 ~\rm{g/cm}^2)$ at $r = 5$ AU.

When $a < 9\lambda/4$, Epstein drag from molecular collisions applies \citep{ahn76}.  Then the particle size for a given (midplane) $\taus$ is
\begin{equation} 
a_{\rm Ep}(\taus) = {\taus \varSigma\gs \over \sqrt{2\pi}\rho_\bullet}  \approx 30~ {\rm cm}~\taus \varSigma_{150} \left(\frac{r}{5 \rm AU}\right)^{-1.5}\, .\label{eq:aEp}
\end{equation} 
For $a > 9\lambda/4$, particles enter the viscous Stokes drag regime and
\begin{equation} \label{eq:aSt}
a_{\rm St}(\taus) = \sqrt{9 \mu c\gs \ts \over 4 \rho_\bullet \sigma_{\rm mol}} = 90 ~{\rm cm}~\sqrt{\taus}  \left(\frac{r}{5 \rm AU}\right)^{5/8}\, ,
\end{equation} 
independent of gas density and with $c\gs$ as the sound speed.  For linear drag the particle size is given by the \emph{smaller} of $a_{\rm Ep}$ or $a_{\rm St}$.

Drag forces become nonlinear when the Reynolds number, ${\rm Re} = \Delta V a/ (\lambda c\gs)$, of the flow around the particle exceeds unity.  We take the relative motion of particles and gas, $\Delta V \sim \eta v_{\rm K} \sim c\gs^2 / v_{\rm K}$, from the pressure supported gas rotation a small fraction of the Keplerian speed, $v_{\rm K} = \varOmega r$, see \Eq{eq:eta}.  We estimate the size and stopping time, from \Eq{eq:aSt}, of the transition to turbulent drag as:
\begin{eqnarray} 
a_{\rm turb} \sim \lambda v_{\rm K}/c\gs &\approx& 25 ~{\rm m}~\varSigma_{150}^{-1} \left(\frac{r}{5 \rm AU}\right)^{2.5} \label{eq:aturb}\\
\tau_{\rm s,turb} \sim \sqrt{2\pi} {4\rho_\bullet \lambda \over 9 \varSigma\gs}\left({v_{\rm K} \over c\gs}\right)^2 &\approx& 700~ \varSigma_{150}^{-2}\left(\frac{r}{5 \rm AU}\right)^{3.75}
\end{eqnarray} 
Our assumption of a linear drag law applies up to large particle sizes in the outer disk, but only for $\taus \lesssim 1$ at $r \lesssim 1$ AU.

\begin{table}[h]
\caption{Symbols}
\begin{center}
\begin{tabular}{|c|c|c|}
\hline
Symbol & Definition/Use & Meaning \\
\hline
$\varOmega$ & $ v_{\rm K}/r$& Keplerian orbital frequency\\
$\ts$ & eqs.\ (\ref{eq:aEp}),(\ref{eq:aSt})  & particle stopping time \\
$\te$ & eq.\ (\ref{eq:power}) & eddy turnover time \\
$t_{\rm sett}$ &eq.\ (\ref{eq:tsett}) & vertical settling time\\
$\taus$ & $\varOmega \ts$ & dimensionless stopping time \\
$\taue$ & $\varOmega \te$ & dimensionless eddy time \\
$\St$ & $\ts/\te$ & Stokes number\\
$\Sc_{(x)}$ & $D\gs/D\ps$, eq.\ (\ref{eq:Scx}) & Schmidt number (radial)\\
$D\gs, D_{\mathrm{p},(x)}$& eqs.\  (\ref{eq:Dg}), (\ref{eq:Dpx})&diffusion coefficients: gas and particle (radial)\\
$\alpha_{(x,z)}$ & $D\gs \Omega/c\gs^2$, eq.\ (\ref{eq:Hpstd})&  
dimensionless diffusion coefficients\\
$H\ps$, $H\gs$ & eq.\ (\ref{eq:Hpsq})& particle, gas scale heights\\
$\delta V\gs, \delta u\gs, \delta w\gs$&Tab.\ \ref{tab:ordering}& RMS turbulent speeds: isotropic, radial, vertical\\
$\delta V\ps, \delta u\ps, \delta w\ps$& eq.\ (\ref{eq:vpest}), \Fig{fig:wpts} & RMS particle speeds: isotropic, radial, vertical\\
$\brak{\ug^2}, \brak{\vg^2},\brak{\wg^2}$ & eqs.\  (\ref{eq:ugsq}),(\ref{eq:turbmodel}) &  $\begin{array}{c}
{\rm mean~ squared~ turbulent~ gas~ speeds:}\\
{\rm radial, azimuthal, vertical}\\
\end{array}$\\
$\brak{\up^2}, \brak{\vp^2},\brak{w\ps^2}$ & eqs.\  (\ref{eq:upsq}),(\ref{eq:episq}),(\ref{eq:wpsq}) & $\begin{array}{c}
{\rm mean~ squared~ random~ particle~ speeds:}\\
{\rm radial, azimuthal, vertical}\\
\end{array}$\\
$ \brak{\ug\vg}, \brak{\up\vp}$& eqs.\ (\ref{eq:turbmodel}), (\ref{eq:episq}c) & velocity correlations: gas, particle\\
$c\gs$ &eqs.\  (\ref{eq:aSt}),  (\ref{eq:aturb})& isothermal sound speed\\
$r$ &eq.\  (\ref{eq:mfp}) & radial distance from star \\
$x$, $y$, $z$ & eqs.\ (\ref{eq:Hills}),(\ref{eq:vertLangevin})&local radial, azimuthal, vertical coordinates\\
$\rho\ps$, $\rho\gs$&  (\ref{eq:feed}) & particle, gas space density\\
$a,\rho_\bullet$&eqs.\ (\ref{eq:aEp}), (\ref{eq:aSt}) &particle radius, internal density\\
$\varSigma$, $\varSigma\gs$&eq.\ (\ref{eq:Sigma}a)&particle, gas surface density\\
$\eta$ & eq.\ (\ref{eq:eta}) & pressure support parameter\\
$\Delta V$& eqs.\ (\ref{eq:vr}), (\ref{eq:vphi})& particle-gas relative velocity\\
\hline
\end{tabular}
\end{center}
\label{default}
\end{table}

\section{Turbulence Model for Arbitrary Eddy Time}\label{s:turbmodel}

To numerically evaluate our results for in-plane particle stirring (\S\ref{s:Hepi}), we relate the radial, azimuthal and correlated turbulent velocities ($\brak{u\gs^2}$, $\brak{v\gs^2}$, and $\brak{u\gs v\gs}$) as a function of $\tau_{\rm e}$.  The essential assumption is that velocities are isotropic, which we show is reasonable even for spatially sheared eddies with $\taue \gg 1$.  The detailed value of $\brak{u\gs v\gs}$ has a weak effect on particle stirring (and even then only near $\taus = 1$), but is included for completeness.  We also argue that particle trapping is weak when eddies survive for only a single turnover time, even when that turnover time is long.

We separately consider the limiting cases of large and small $\tau_{\rm e}$.  
For $\tau_{\rm e} \ll 1$ we take the unsheared turbulence to be isotropic, $\brak{u\gs^2} = \brak{v\gs^2}$.  The velocity correlations are weak, and generated by the linear shear of radial velocity into azimuthal (which builds up for $\te$) as $\brak{u\gs v\gs} \sim (3/2)\tau_{\rm e}\brak{u\gs^2}$.

The $\tau_{\rm e} \gg 1$ case is more constrained by the dynamics of background shear.
We introduce $\omega'$ as the perturbed vorticity of the turbulence
on the outer scale, which we compare to the background shear vorticity, $\omega_{\rm K} = -3\varOmega/2$. 
(The ratio $|\omega'/\omega_{\rm K}|$ is comparable to the ratio of turbulent speed to sound speed
when the radial outer scale of the turbulence is comparable to the scale height.)
We consider  $|\omega'/\omega_{\rm K}| < 1$, because the efficiency of converting
vorticity from the background shear into turbulence is likely to be less than 100\%.  (Note however that $|\omega'/\omega_{\rm K}| > 1$ is allowed for the forced  $\tau_{\rm e} \ll 1$ turbulence above.)

To build a model for $|\omega'/\omega_{\rm K}| < 1$, we are guided by 
the dynamics of two-dimensional vortices, which have the following well-known properties \citep{saf92, chav00}: (1) vortices exist in regions where the perturbed vorticity $\omega' < 0 $
is more negative than the background vorticity;
(2) vortices are elongated in
the azimuthal direction by a factor $\Delta y/\Delta x \sim |\omega_{\rm K}/\omega'|$ relative to their radial extent; 
(3) the radial and azimuthal (with Kepler shear subtracted) fluid velocities are $u\gs\sim |\omega'^2/\omega_{\rm K}| y$ and  $v\gs \sim - |\omega'| x$.  Thus at the vortex edge $\brak{u\gs^2}^{1/2} \sim \brak{v\gs^2}^{1/2} \sim |\omega'|\Delta x$ and the velocities have the same magnitude;
 (4) the time for fluid
 to circulate around a vortex is $\sim 1/|\omega'|$.
 
We now depart from our analogy with 
long-lived two dimensional vortices which are  are absolutely
stable in the absence of viscosity and are clearly not turbulence---since energy dissipation and angular momentum transport are absent.  In three dimensions, though, vortices decay \citep{bm05, ssg06}.  We hypothesize that three-dimensional turbulence is characterized by the properties (2) and (3) of vortices listed above; and that, based on property (4), 
vortices live for a time $\sim 1/|\omega'|$ before decaying.  Thus we assume $\te \approx 1/|\omega'|$ so the assumption $|\omega'/\omega_{\rm K}| \ll 1$ is equivalent to $\taue \gg 1$.

To calculate the $\langle u\gs v\gs\rangle$ for $\taue \gg 1$ we use the established relation 
that the energy dissipation rate  is $3\varOmega/2$ times the angular momentum flux 
\citep{lp74,lc07}.  (See \citealp{yj07} for how particle feedback introduces pressure work to the energy balance.)
Since the energy dissipation
rate is $\sim \langle u\gs^2+v\gs^2\rangle/t_{\rm eddy}$ and the angular momentum flux is $\langle u\gs v\gs\rangle$, we have $\langle u\gs v\gs \rangle\sim 2 \langle u\gs^2+v\gs^2\rangle/(3\tau_e)$.

Combining the results for the large and small $\taue$ regimes gives 
\begin{equation} \label{eq:turbmodel}
 \brak{\ug^2} = \brak{\vg^2} = {2+9\taue^2/4 \over 3 \tau_{\rm e}} \brak{\ug \vg}\, .
\end{equation} 
The numerical coefficients should not be taken literally, but are for concreteness.  The $\brak{u\gs v\gs}$ correlations given by \Eq{eq:turbmodel} are a reasonable upper limit.  In any event $\brak{u\gs v\gs}$  has only a moderate effect on particle stirring and only near $\taus = 1$.   We also note that with an $\alpha$-parametrization, $\brak{\ug\vg} \equiv \alpha c\gs^2$, the standard approximation $V_{\rm turb} =  \sqrt{\brak{u\gs^2 + v\gs^2} } \approx \sqrt{\alpha} c\gs$ applies for hydro-turbulence only if correlations are strong  $\brak{\ug\vg} \approx \brak{u\gs^2 + v\gs^2}$, i.e.\ near $\taue = 1$ in \Eq{eq:turbmodel}.  In MRI turbulence on the other hand, 
$\alpha$ includes a dominant contribution from 
Maxwell stresses, so $V_{\rm turb} \approx \sqrt{\alpha} c\gs$ can hold with weaker correlations.

Long-lived two-dimensional vortices can concentrate particles \citep{fmb01}.   However since we assume that eddies only survive for a single turnover time, particle concentration is weak, and our diffusive treatment should be a good approximation.
Because the drift speed of a particle towards the center of a vortex is\footnote{
We neglect the extra drift caused by cause drag with the mean shear of the disk.
This is appropriate if $|\omega'/\varOmega|> H/r$, because then the non-Keplerian
velocities due to the vortex are larger than those due to the mean shear.  If this
inequality were violated, then particle would 
drift straight through vortices and concentration would be negligible.}
 $u_{\rm drift} \sim v\gs \tau_s/(1+\tau_s^2)$,  the time for particles
to reach the center of a vortex is $\Delta x/u_{\rm drift} \sim (\Delta x /v\gs)(1+\tau_s^2)/\tau_s\sim t_{\rm eddy}
(1+\tau_s^2)/\tau_s$, i.e., it is always longer than $t_{\rm eddy}$.  Vortices must survive for many circulation times for particle concentration to be
important.

\citet{kh97} consider particle concentration by a ``sideways'' vortex, i.e.,
a vortex whose
rotation axis is parallel to the midplane.  Such a vortex    concentrates
particles in  the particles' settling time if $|\omega|\lesssim \varOmega$.
\citet{kh97} show that sideways vortices  appear in 2D
simulations of convection,
so the efficiency of this mechanism in 3D with radial drift deserves more study.

\bibliography{}\

\end{document}